\documentclass[prb,twocolumn,showpacs,floatfix,superscriptaddress]{revtex4}
\usepackage{amsmath,amsfonts}
\usepackage{graphicx}
\usepackage{color}
\usepackage{enumerate}
\usepackage{mathtools}
\usepackage{bbm}
\usepackage{wrapfig}
\usepackage{bm}
\usepackage{hyperref}
\usepackage{esint}

\newcommand{\e}{\mathrm{e}}

\newcommand{\bsigma}{\boldsymbol{\sigma}}
\newcommand{\ua}{\uparrow}
\newcommand{\da}{\downarrow}

\newcommand{\s}{\sigma}

\newcommand{\bS}{\mathbf{S}}

\newcommand{\br}{\mathbf{r}}
\newcommand{\bk}{\mathbf{k}}
\newcommand{\bp}{\mathbf{p}}

\newcommand{\ket}[1]{| #1 \rangle}
\newcommand{\bra}[1]{\langle #1 |}

\newcommand{\beq}{\begin{equation}} 
\newcommand{\eeq}{\end{equation}}

\newcommand{\nn}{\nonumber}

\begin{document}



\title{Topological properties of chains of magnetic impurities on a superconducting substrate: Interplay between the Shiba band and ferromagnetic wire limits}

\author{Gian Marcello Andolina}
\email{gian.andolina@sns.it}
\affiliation{Laboratoire de Physique des Solides, CNRS,
             Univ. Paris-Sud,
Universit\'e Paris Saclay, 91405 Orsay cedex, France}
\affiliation{NEST, Scuola Normale Superiore, I-56126 Pisa, Italy}
\affiliation{Istituto Italiano di Tecnologia, Graphene Labs, Via Morego 30, I-16163 Genova, Italy}

\author{Pascal Simon}
\affiliation{Laboratoire de Physique des Solides, CNRS,
             Univ. Paris-Sud,
Universit\'e Paris Saclay, 91405 Orsay cedex, France}

\date{\today}


\pacs{
74.20.Mn, 
71.10.Pm, 
75.30.Hx, 
75.75.-c  
}



\begin{abstract}

We consider a one-dimensional system combining local magnetic moments and a delocalized metallic band on top of a superconducting substrate. This system can describe a chain of magnetic impurities with both localized polarized orbitals and delocalized s-like orbitals or a conducting wire with embedded magnetic impurities. We study the interplay between the one-dimensional Shiba band physics arising from the interplay between  magnetic moments and the substrate and the delocalized wire-like conduction band on top of the superconductor. 
We derive an effective low-energy Hamiltonian in terms of two coupled asymmetric Kitaev-like Hamiltonians  and analyze its topological properties. We have found that this system can host multiple Majorana bound states at its extremities provided a magnetic mirror symmetry is present.
We compute the phase diagram of the system depending on the magnetic exchange interactions, the impurity distance and especially the coupling between both bands. In presence  of inhomogeneities which typically break this magnetic mirror symmetry, we show that the coexistence of a 
 Shiba and wire delocalized topological band
can drive the system into a non-topological regime with a splitting of Majorana bound states.
\end{abstract}


\maketitle


\section{Introduction}
Topological superconductors have received much attention recently,
partly because they host exotic low energy excitations such as Majorana bound states (MBS), \cite{Kane2010,Leijnse2012,Beenakker2013}
whose non-Abelian statistics are attractive for topological quantum computation. \cite{Nayak2008,Pachos2012}
Several different platforms to realize topological superconductivity are currently the subject of intensive research. 

A rather simple recipe combining arrays of magnetic atoms or nanoparticles on top of a  superconducting surface has attracted attention in the past years.\cite{Choy2011,Nakosai2013,NP2013,Braunecker2013,Klinovaja2013,Vazifeh2013,Pientka2013,Pientka2014,Poyhonen2014,Reis2014,Kim2014,Li2014,Heimes2014,Brydon2015,Ojanen2015a,Peng2015,Rontynen2015,Hui2015,Braunecker2015,Poyhonen2016,Zhang2016,Li2016a,Rontynen2016,Hoffman2016,Li2016b,Kaladzhyan2016b,Schecter2016,Christen2016,Kalad2017}
Recent experiments on chain of iron atoms  adsorbed on lead have been realized experimentally  and 
revealed the existence of zero bias peaks spatially localized on the end of such chain which have been interpreted as signatures of Majorana bound states.\cite{NP2014,Pawlak2016,Ruby2015,Yazdani2017} Instead, Cobalt atomic chains adsorbed on lead seem not to give rise to a topological phase hosting protected MBS.\cite{Ruby2017}
In order to describe these  experiments, at least two different types of models have been used. 

In the first model corresponding to the dilute impurity regime, we can either assume that the  magnetic atoms can be described by classical isolated spins which induce Shiba bound states\cite{Yu1965,Shiba1968,Rusinov1969,Balatsky2006} in the superconducting substrate. The overlap between these Shiba bound states entails the formation of  a one-dimensional (1D) Shiba band inside the superconductor which may eventually be in a topological phase provided some conditions are met.\cite{NP2013,Pientka2013,Pientka2014,Poyhonen2014,Reis2014} In this description, the magnetic atoms are assumed to be fully polarized and their orbitals have a negligible overlap which corresponds to the dilute limit. Furthermore, some magnetic texture, typically a planar helix, is  {\it a priori} assumed to take place before hand. Such magnetic texture  could come from the combination of RKKY interactions, crystal field, and spin-orbit coupling. This limit seems however  not to correspond to the experiments where a ferromagnetic dense iron wire is deposited on the superconducting lead substrate.\cite{NP2014,Pawlak2016,Ruby2015,Yazdani2017} 

Alternatively, in the second model corresponding to the dense magnetic impurity limit, 
 the major role of the superconducting substrate seems to induce a proximity induced gap in the ferromagnetic wire.\cite{NP2014,Li2014,Kim2014,Yazdani2017} Such description is actually closer in spirit to recent experiments performed with semiconducting  wires in proximity of a bulk superconductor \cite{Mourik2012} or epitaxially grown semiconductor-superconductor nanowires.\cite{Krogstrup2015,Albrecht2016,Deng2016,Kouwenhoven2016} When the ground state of the isolated wire is ferromagnetic, an effective spin texture necessary to enter into a topological phase is brought by the combination of the exchange field and the spin-orbit coupling which can be either intrinsic to the wire\cite{NP2014,Li2014,Yazdani2017} or extrinsically brought by the substrate.\cite{Kim2014} Note also that such helical field can also come from a self-tuning RKKY interaction mediated  by the 1D wire conduction electrons between the magnetic spins.\cite{Braunecker2013,Klinovaja2013,Vazifeh2013,Braunecker2015}

In the dilute Shiba chain limit, the iron atoms are treated as effective classical magnetic fields to create bound states inside the superconductor. When the Shiba band is topological, the MBS are localized inside the substrate. However, in the  wire limit, the superconducting substrate can be  integrated out and the system becomes analogous to a conducting wire with local exchange magnetic field proximitized by a superconductor. In the topological phase, the MBS are mainly formed within the 1D wire  conduction band. The  picture emerging from these two limiting cases are thus qualitatively different. One can go from one regime to the other by modeling this sytem as a linear chain of Anderson impurities with a non-zero hybridization between the atoms.\cite{Peng2015}

By simply superposing the previous two limits, one may naively expect to find at least two types of MBS,  localized either in the substrate
or in the wire. However, from the point of view of the fundamental symmetries taking place here, this system is particle-hole symmetric and breaks time reversal symmetry (TRS). The chain of magnetic atoms is thus expected to be in class D and characterized by a ${\mathbb Z}_2$ invariant.\cite{Kitaev2009,Ryu2010} It should  therefore host at most one MBS at the extremity of the chain 
except if some low energy chiral-like symmetry is emerging driving the system in the BDI class.

However, one may wonder in which experimental systems and cases  both the Shiba band and the wire band shall be taken into account. 
The following systems could be envisioned: consider an array of magnetic impurities whose distance can be controlled at the atomic level using tip manipulation. One may thus depart from the dense limit considered in [\onlinecite{NP2014,Pawlak2016,Ruby2015,Yazdani2017}] and explore an intermediate distance regime. Such strategy is presently followed in [\onlinecite{Wiesendanger}].
If the impurities have  both localized polarized d-like orbitals and delocalized s-like orbitals adsorbed on a superconducting substrate, then the overlap between the impurities  and their interaction with the substrate shall be taken into account and the theory developped in this paper may apply. 
This scenario  can also apply to a 1D conducting structure with embedded magnetic moments deposited on a superconductor. This could be the case use of a 1D  assembly of magnetic molecules on top of the  superconductor surface.  Supramolecular chemistry and self-assembly concepts are fast developing techniques that could be utilized to create atomically defined systems with controlled and tunable interaction between  periodically spaced magnetic centers. Potential candidates are  porphyrin-based molecular nanowires
\cite{Zheng2016} or  Mn-based metal–organic networks \cite{Giovanelli2014} to list only a few. In these kinds of systems, magnetic atoms interact both with the substrate but also with each others via the organic molecules (and also via the substrate). Therefore, such systems may also offer an interesting platform where Shiba bands could coexist with a 1D conduction band on top of a superconducting substrate.

In this paper, we consider such intermediate situation where localized magnetic moments interact with both  a two-dimensional (2D) superconducting substrate and a 1D delocalized conduction band. 
In the deep-Shiba limit, we obtain a low-energy 4-band Hamiltonian describing coupled Shiba and wire bands. Under some conditions, such as magnetic moments forming a perfect planar helical texture and no other source of  inhomogeneities, we have found  that this low-energy 4-band Hamiltonian has an effective time-reversal symmetry
 which casts it in the BDI class able therefore  to sustain multiple Majorana fermions at the extremities of the chain. We have shown that this effective TRS can be traced back to a magnetic mirror symmetry,\cite{Bernevig2014} akin to a crystalline symmetry in topological insulators,\cite{Fu2011,Ando2015} which protects these multiple Majorana edge modes  from hybridizing as found in the wire impurity description.\cite{NP2014,Li2014}
We have  computed the complex phase diagrams of the system depending on the magnetic exchange interactions, the impurity distance and especially the coupling between the Shiba and wire bands.
When this fragile  magnetic mirror symmetry is broken (thus the effective TRS), the phase diagrams simplify drastically. In particular, phases with two MBS become topologically trivial  suggesting that there is an intermediate regime between the dilute and dense limit where MBS do hybridize.
 
The plan of the paper is as follows: In Sec. 2, we describe our generic model Hamiltonian to take into account both the Shiba and 1D wire delocalized bands and discuss various limiting cases that recover well established results in the literature. In Sec. 3, we derive our effective low-energy 4-band Hamiltonian and discuss its symmetry properties with emphasis on this magnetic mirror symmetry. In Sec. 4, we derive topological phase diagrams of this system depending whether the magnetic mirror symmetry is present or not. Finally, in Sec. 5 we summarize and discuss  our results.


\section{Description of the system}
\subsection{Model Hamiltonian}
\begin{figure}[h]
	\centering
	\includegraphics[width=1\linewidth]{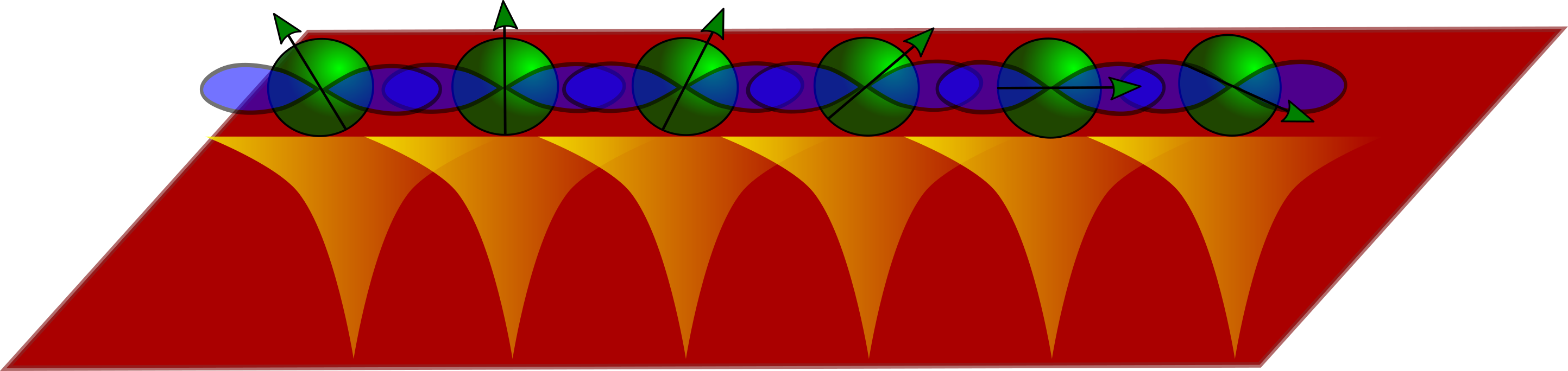}

	\caption{ {A sketch of the system studied in
this work. The system is made by an array of magnetic adatoms (in green) deposited on a 2D s-wave superconducting substrate (in red). The magnetic interaction $J$ between the magnetic adatom and the substrate induces a localized Shiba bound state (in yellow) in the substrate. An array of magnetic adatoms creates a Shiba band below the superconducting band. Furthermore, we also take into account delocalized orbitals (in blue), interacting  with the magnetic adatom with an exchange interaction $J^\prime$. An array of such adatoms creates a conducting band made of delocalized electrons. }
	}
	\label{sketch}
\end{figure}
We consider an array of magnetic impurity on a superconducting substrate as depicted in Fig. \ref{sketch}.
Our starting point is the Hamiltonian of a 2D s-wave superconductor in the clean limit described by the following Hamiltonian
\begin{align}
H_{\rm SC}=&\int d^2 p \sum_\s \xi_p\psi^\dag_\s(\bp)  \psi_\s(\bp) \nn \\
&+\int d^2 r [\Delta \psi_\ua^\dag(\br)\psi_\da^\dag(\br)+h.c.].
\end{align}
Here $\bp$ and $\br$ denote the the electron’s momentum and position,
$\xi_p=p^2/2m-\mu$ is the dispersion relation of the substrate with $\mu$ the chemical potential.  The parameter $\Delta$ is the s-wave superconducting gap. The electron field $\psi_\ua^\dag(\br)$ creates an electron of spin $\ua$ at the position $\br$ in the substrate.
Although our analysis can be straightforwardly extended to a 3D substrate, we consider a 2D one because it has been shown experimentally that the Shiba in-gap states in a 2D superconductor have a much larger spatial extent that its 3D counterpart\cite{Menard2015} which facilitates the formation of Shiba bands.
We suppose that the superconductor hosts an array of
magnetic impurities placed at position $\br_n$. We are interested in a regime able to interpolate between the ferromagnetic wire placed on top of the substrate where the impurities have a strong orbital overlap and the regime of dilute magnetic impurities with negligible orbital overlap. In order to cover these two regimes, at least phenomenologically, we assume that each magnetic impurity has at least two  orbital degrees of
freedom:  some well localized d-like  orbitals which are polarized and can be 
described  with a classical spin $S$ and a delocalized s-like conduction electron orbital characterized by the operator $a^\dag_{n\s}$ creating an electron with spin $\s$ at position $\br_n$. 
We would like also to point that such system could also describe a conducting wire with embedded magnetic atoms adsorbed on a superconducting substrate.
We assume that the magnetic moments, provided by the
polarized localized electrons can be described a  classical spin  $S\gg 1$ and arranged along a linear chain 
with lattice spacing $|\br_n-\br_{n+1}|=a$. We restrict ourselves to the dilute limit with typically $k_F a\gg 1$ with $k_F$ the Fermi momentum of the substrate.
Furthermore, we assume that we can parametrize the impurity spins 
$\bS_n$ through spherical coordinates, supposing a planar helical order, characterized by the helical momentum $k_h$ such as
\begin{equation}
\bS_n=
\begin{pmatrix}
\cos(2\phi_n) ,&\sin(2\phi_n), & 0
\end{pmatrix},
\end{equation}
with $\phi_n=k_hna$.
We do not  focus on how this kind of given spin texture arises. Indeed, such spin arrangement can be obtained either intrinsically if the magnetic impurities interact mainly via indirect interactions along the 1D conducting channel \cite{Braunecker2013,Klinovaja2013,Vazifeh2013}
or effectively after a unitary transform to gauge away a strong spin-orbit interaction in the superconducting substrate. \cite{Kim2014,Zhang2016}

The array of magnetic impurities forming a ferromagnetic wire is thus described by the Hamiltonian $H_{\rm wire} + H_{\rm imp}$ with
\begin{align}
H_{\rm wire} 
&= \sum\limits_n\sum\limits_{\alpha} \left[-t_w( a^\dag_{n+1,\alpha} a_{n,\alpha} + h.c.)
+\mu_w a^\dag_{n,\alpha} a_{n,\alpha} \right], \nonumber\\
H_{\rm imp}&=\sum\limits_n\sum\limits_{\alpha,\beta}\left[ -J\psi_\alpha^\dag(\br_n)\bS_n\cdot\bsigma_{\alpha\beta} \psi_\beta(\br_n)\right.\nonumber\\ &\hskip 1cm
\left. -J' a^\dag_{n,\alpha} \bS_n\cdot\bsigma_{\alpha\beta} a_{n,\beta}\right].
\end{align}
Here, $a^\dag_{n,\alpha}$ creates an electron at position $\br_n$ with spin $\alpha$ in the wire,
$t_w$ is the wire hopping parameter between the delocalized orbitals of two neighboring impurities, and $\mu_w$ the  chemical potential of this conduction band which we assume positive for further convenience. The couplings $J$ (respectively $J'$) denotes the magnetic exchange interaction between the spin $\bS$ and the electrons of the superconducting substrate (respectively the conduction electrons of the delocalized 1D band).
The  wire conduction band is also coupled to the superconductor via a tunneling Hamiltonian
\begin{equation} \label{eq:htun}
H_{\rm tun}=-t\sum\limits_{n,\alpha} \left[\psi_\alpha^\dag(\br_n) a_{n,\alpha}+h.c.)\right].
\end{equation}
The parameter $t$ describes the hopping between the wire delocalized electrons and the substrate. 
In the experiments of iron magnetic wire on top of lead, \cite{NP2014,Pawlak2016,Ruby2015} the magnetic moments order ferromagnetically. In order to obtain some topological superconductivity, it is then necessary to include some Rashba spin-orbit coupling in the superconducting substrate.\cite{NP2014,NP2013,Kim2014} 
We note that such helical order can be mapped to a ferromagnetic order but with some spin-orbit coupling in both the superconducting substrate and the conduction band of the wire electrons after some unitary transform. \cite{Braunecker2011,Kim2014} The total Hamiltonian of the system is thus
 \beq \label{eq:H}
H=H_{\rm SC}+H_{\rm wire}+H_{\rm imp} + H_{\rm tun}.\eeq
Although $H$ is quadratic, it contains many energy scales and uncovers many different physical situations. Therefore we will first  consider various limiting cases.

\subsection{Qualitative analysis}
As stressed in the introduction, the interesting situation we want to describe corresponds to the case where both the 1D conduction channel and the 1D Shiba band are present at low energy. Before describing such novel situation, let us first discuss some well-known limits.

\subsubsection{The $J'\to 0$ limit}
In Eq. \eqref{eq:H}, we introduce $J$ and $J'$ which are two magnetic exchange interactions. When $J'=0$, we are left with classical impurities interacting only with the SC substrate. The conducting wire electrons can be integrated out and we are left with a SC  substrate with some slightly renormalized value of its parameters. 

An isolated classical magnetic  impurity exchanged coupled with the s-wave superconductor produces a  Yu-Shiba-Rusinov bound state\cite{Yu1965,Shiba1968,Rusinov1969,Balatsky2006} (called Shiba in what follows) of energy 
\beq\epsilon_S=\Delta\frac{1-\alpha^2}{1+\alpha^2},\eeq 
with $\alpha=\pi\nu JS$,
 where $\nu_0$ is the density of states of the substrate. 

The array of impurities gives rise to a Shiba band which may turn to be topological and host up to  two Majorana fermions below the extremities of the chain. This situation has been extensively studied previously [\onlinecite{Pientka2013,Pientka2014}] and we will not detail it here. In such case, the Majorana wave functions builts in the substrate at both extremities of the chain. We expect this case to be an appropriate description as soon as $J'\ll J$.

\subsubsection{The $J\to 0$ limit}
Let us now describe the other opposite limit. This corresponds to an array of magnetic moments embedded in a 1D conducting channel proximitized to a bulk superconductor. Such a situation has been extensively treated in Refs \onlinecite{Braunecker2013,Klinovaja2013,Vazifeh2013,Braunecker2015}. The conduction band mediates a 1D RKKY interaction between the impurity spins which favors a spiral alignment of the magnetic moments. Using a self-consistent calculation it was established that the topological phase self-tunes without any adjustable parameters.\cite{Braunecker2013,Klinovaja2013,Vazifeh2013} This means that the Majorana phase is the ground state of this 1D proximitized conducting channel (provided the magnetic exchange energy is larger than the proximity induced gap). Here the Majorana bound states are mostly localized at the 1D wire conduction band. We expect this limit to hold while $J\ll J'$.

\subsubsection{Comparable magnetic exchange $J\sim J'$}
From the previous discussion, an interesting situation may occur when the two magnetic exchange couplings are comparable. One may expect an interplay between the 1D Shiba band  and the 1D proximitized conduction band.
 This corresponds to the situation where both 1D channels can eventually coexist at  low energy near the middle of the superconducting gap. This means  $\epsilon_S\ll \Delta$ and corresponds to the so-called deep shiba limit. This implies therefore a very strong magnetic exchange energy scale $J\sim J'\gg \Delta$.
Another important parameter in the system is the distance $a$ between the impurities. 
As mentioned above, we consider the rather dilute impurity limit $k_F a\gg 1$. We also assume $a\ll \xi$ where $\xi$ denotes the superconducting coherence length. This limit is  met is all experiments working with arrays of magnetic adatoms adsorbed on a 3D substrate.\cite{NP2014,Pawlak2016,Ruby2015} Two scenarios can be envisioned: either several MBS can coexist eventually  protected by some low energy emerging symmetry or there is a strong hybridization between the MBS which splits them away from zero energy. This is exactly what we are going to show.

\section{Low-energy effective Hamiltonian and symmetry considerations}
\subsection{Derivation}
In this section, we derive a low-energy Hamiltonian for the case of main interest in this work with comparable exchange energy scales $J\sim J'$. 
Following [\onlinecite{Pientka2013}], the dilute impurity limit $k_F a \gg 1$ guaranties that the Shiba and wire conduction bands are within the superconducting gap $\Delta$. Our strategy is thus to integrate out high energy degrees of freedom of energy $|E|\geq \Delta$ to obtain a low-energy effective Hamiltonian for both the Shiba band and the wire conduction band.

Let us denote $\ket{n}$  the Shiba state associated with a magnetic impurity placed in the site $n$.
An effective Hamiltonian for the Shiba chain can be obtained by projecting on the single impurity Shiba  states 
following.\cite{Pientka2013} For an array of $N$ magnetic impurities, we obtain a $2N\times 2N$ tight binding Bogoliubov-de Gennes  (BdG) Hamiltonian.
After the projection we obtain a BdG effective Hamiltonian $H_\text{Shiba}^{\rm eff}$ for the Shiba chain

\begin{eqnarray}
\label{eq:HShiba}
\left(H_\text{Shiba}^{\rm eff}\right)_{mn}&=&\bra{m} H_\text{Shiba}^{\rm eff}\ket{n}\nn\\
&=&
 \begin{pmatrix}    ( h^\text{eff})_{m,n} &  ( \Delta^\text{eff})_{m,n} \\
 ( \Delta^\text{eff})_{m,n} & -( (h^T)^\text{eff})_{m,n}\end{pmatrix}~,
\end{eqnarray}
with
\begin{eqnarray}
\label{H_effShiba}
 ( h^{\text{eff}})_{m,n}&\approx& \Delta(1-\alpha) \delta_{m,n}+\cos(k_F r_{mn}-\frac{\pi}{4}) \nn\\ 
&\times&  \cos(\bk_h \cdot \br_{mn}) f(r_{mn})(1-\delta_{m,n}), \\
 ( \Delta^\text{eff})_{m,n}&\approx&i \sin(k_F r_{mn}-\frac{\pi}{4}) \nn\\
 &\times& \sin(\bk_h \cdot \br_{mn}) f(r_{mn})(1-\delta_{m,n}),
\end{eqnarray}
where $f(r)=-\Delta\sqrt{\frac{2}{\pi k_Fr }}e^{-r/\xi}$.
 Here $r_{mn}=|\br_{mn}| = |\br_m-\br_n|$ is the distance between two impurity lattice sites.
We have assumed above that $k_F a \gg 1$, which guarantees a small Shiba bandwidth and allowed us to use an approximate long range analytical expansion.

Now we focus on  the delocalized wire electrons.
We previously projected out the  states of the  superconducting substrate with energy $|E|>\Delta$. This will affect the $a$-electrons as well and provide
a self-energy of the form  $\Sigma(E)_{m,n}=t^2 \mathcal{G}^{\text{SC}}_{m,n}(E) $, where $\mathcal{G}^{\text{SC}}_{m,n} $ is the exact propagator of the substrate. Here we assumed $t\nu_0\ll 1$ with $\nu_0$ the  density of states in the host superconductor. Because the magnetic impurities do not affect in a significant manner high energy states, we can approximate this propagator by the bare one $ \mathcal{G}^{\text{SC}}_{m,n} \approx \mathcal{G}^{0}_{m,n}$
so that  $\Sigma(E)_{m,n}=-\pi t^2 \nu_0\frac{E+\Delta \tau_x}{\sqrt{\Delta^2-E^2}}$.
At  low energies $E\ll\Delta$, we thus obtain a proximitized pairing  term for the $a$-electrons of the form
$-\Delta'\sum_{n}(a_{\uparrow,n}^\dagger a^\dagger_{\downarrow,n}+h.c.)$,
with $\Delta'=t^2\pi\nu_0$.

In the large magnetization case we are interested in, $J'S\sim JS \gg t_w$ which implies that the bands of the ferromagnetic wire are well separated energetically and polarized.
We first perform a local unitary transform $U_n$ to locally align the conduction electron spin along   $\hat{\vec{S}}_n$ such that $\tilde{a}^\dagger_{\alpha ,n}= (U_n)_{\alpha,\beta} {a}^\dagger_{\beta,n}$ with
\beq
U_n=\frac{1}{\sqrt{2}}\left(\begin{array}{cc} 
e^{\frac{-i\phi_n}{2}} & e^{\frac{-i\phi_n}{2}}\\e^{\frac{i\phi_n}{2}}& -e^{\frac{i\phi_n}{2}}\end{array}\right).
\eeq
Non trivial physics occurs only when  the Shiba band and the polarized conduction band are both at the Fermi energy. We therefore assume $\mu_w\sim J'S\gg t_w$. If this is not the case, we are  left with only the Shiba band at low energy, a situation treated in details in [\onlinecite{Pientka2013,Pientka2014}].
We then project on the lower polarized conduction band paying attention to possible virtual processes occurring in the upper conduction band. 
Following [\onlinecite{Choy2011}], we obtain

\begin{eqnarray}\label{eq:Hwire}
&&H^{\text{eff}}_{\text{wire}}\approx \sum_{n} \tilde{a}_{n,\uparrow}^\dagger  \biggl(     \mu_w-J'S -\delta \mu_w\biggr)\tilde{a}_{n,\uparrow} \\ &+&
\sum_n[- \frac{t_w}{2}\cos(k_ha)  \tilde{a}_{n,\uparrow}^\dagger \tilde{a}_{n+1,\uparrow}+\delta t_w  \tilde{a}_{n,\uparrow}^\dagger \tilde{a}_{n+2,\uparrow}+h.c.] \nn \\
&-&\biggr(\frac{1}{J'S}+\frac{1}{\mu_w}\biggl)\sum_n[\frac{\Delta'it_w\sin(k_ha)}{4} \tilde{a}_{n,\uparrow}^\dagger \tilde{a}_{n+1,\uparrow}^\dagger+h.c.]\, , \nn
\end{eqnarray}
where $\delta \mu_w=\frac{t_w^2\cos^2(k_ha)}{4J^\prime S}$ and $\delta t_w=\frac{t_w^2\sin^2(k_ha)}{4J^\prime S}$ are negligible terms that simply renormalize  the chemical potential and add a next to nearest hopping term respectively.
We have thus obtained two 1D spinless bands. However, they are not independent since the electronic degrees of freedom are initially directly  coupled via the tunneling Hamiltonian in \eqref{eq:htun}. 
The coupling term is obtained by projecting
the tunneling Hamiltonian \eqref{eq:htun} onto these two 1D bands to obtain the full low-energy Hamiltonian. 
In order to write this term in the BdG formalism we define $\bra{n,\uparrow_w}$ the lower polarized band  of the wire at the site $n$.
Projecting the  tunneling term, we obtain

\begin{equation}
\label{eq:tunneling}
\begin{aligned}
(H^{\rm eff}_{\rm tun})_{nn}&=\bra{n,\uparrow_w} {H}_\text{tun} \ket{n}= - t\sqrt{\frac{\Delta}{2JS}     }
\begin{pmatrix}
1 & 0 \\
0& -1
\end{pmatrix},\\
(H^{\rm eff}_{\rm tun})_{m\ne n}&=\bra{m,\uparrow_w} H_\text{tun} \ket{n}_{m \neq n}\\
&= -t\sqrt{\frac{\Delta}{2JS\pi k_Fr}   } M(\br_{mn}),
\end{aligned}
\end{equation}
where
$$
M(\br)=
\begin{pmatrix}
\cos(k_Fr) \cos(\bk_h\cdot \br) & i \sin(k_Fr)\sin(\bk_h\cdot \br )\\
i\sin(k_Fr)\sin(\bk_h\cdot \br )& - \cos(k_Fr)\cos(\bk_h\cdot \br )
\end{pmatrix}.
$$
The projection of the tunneling Hamiltonian provides two terms: a diagonal one where an electron  tunnels from one band to the other at the same site $n$,  and a non-diagonal one due to the long range extent of the Shiba wave function. 
Gathering all terms, the resulting effective low-energy tight binding Hamiltonian of our system thus reads 

\beq \label{eq:heff}
H^{\rm eff}=
\begin{pmatrix}
H^{\text{eff}}_{\text{wire}}&H^{\rm eff}_{\rm tun}\\
 (H^{\rm eff}_{\rm tun})^\dagger &H^{\text{eff}}_{\text{Shiba}}
\end{pmatrix},\eeq
where $H^{\text{eff}}_{\text{Shiba}}$ and $H^{\rm eff}_{\rm tun}$ are given  in Eq. (\ref{eq:HShiba}) and Eq. (\ref{eq:tunneling}) respectively. Using Eq. (\ref{eq:Hwire}), $H_\text{wire}^{\rm eff}$ reads

 \begin{equation}
\begin{aligned}
\left(H_\text{wire}^{\rm eff}\right)_{mn}=&\bra{m,\uparrow_w} H_\text{wire}^{\rm eff}\ket{n,\uparrow_w} \\
=&
 \begin{pmatrix}    ( h_{w})_{m,n} &  ( \Delta_{w})_{m,n} \\
 ( \Delta_{w})_{m,n} & -( (h_{w})^T)_{m,n}\end{pmatrix},
\end{aligned}
\end{equation}
with \beq (h_{{w}})_{m,n}= (\mu_w-J'S )\delta_{m,n}-\frac{t_w}{2}\cos(k_ha)\delta_{n,m+1},\eeq
 and  
\beq (\Delta_{w})_{m,n}=
(\frac{1}{J'S}+\frac{1}{\mu_w})\frac{it_wt^2\pi\nu_0\sin(k_ha)}{4}\delta_{n,m+1}.\eeq

$H^{\rm eff}$ in Eq. \eqref{eq:heff} thus describes  two Kitaev-like Hamiltonians coupled with some long-range tunneling terms. 
Before analyzing the phase diagram associated the topological properties of $H^{\rm eff}$, we discuss the symmetry properties of the low-energy Hamiltonian with respect to the symmetry properties of the initial system.

\subsection{Symmetry analysis}
 The Hamiltonian in Eq. \eqref{eq:heff} being of Bogoliubov-De Gennes type,  is invariant under particle-hole symmetry (PHS) by construction.
 $H^{\rm eff}$ is made of spinless fermions. Therefore, the 
time reversal symmetry (TRS) operator simply reads $\mathcal{T_{\rm eff}}=K$, where $K$ is the complex conjugation ($\mathcal{T_{\rm eff}}^2=1$). We emphasize that $\mathcal{T}$ is only an effective low-energy TRS operator not to be confused with $T$, the TRS operator of our initial electronic system.
This is worth noting that the Hamiltonian  in Eq. (\ref{eq:heff}) can be made real with the unitary transform $a_i\longrightarrow e^{i\pi/4}a_i$ and $\Psi_i\longrightarrow e^{i\pi/4}\Psi_i$.
Therefore,  $H^{\rm eff}$ being invariant  under  both TRS  and  PHS  falls into the BDI class with a $\mathbb{Z}$ topological invariant.\cite{Ryu2010,Kitaev2009} $H^{\rm eff}$ can thus  sustain an integer number of Majorana bound states at the extremities of the chain.

As already noticed before for a ferromagnetic wire on top of a superconducting substrate in [\onlinecite{Bernevig2014,Li2014}] such effective TRS of the Hamiltonian is in fact connected to the magnetic group symmetry $M_T$ of the initial system (see appendix \ref{MMS} where we detail this connection).
Therefore, in order  for the system to sustain more than one MBSs at one extremity,
we need it to be invariant under $M_T$. This implies a perfectly aligned chain and a planar spin helix.
Furthermore, the substrate needs to be free of disorder to respect ${\cal M}(y\to -y)$, at least on average.

One can reach the same conclusion by inspecting  how  $H^{\rm eff}$ was obtained. 
Any   complex term which cannot be gauged away breaks the effective TRS. This will be the case if we the helix is no longer planar.\cite{Pientka2013,Pientka2014} This can also be the case if the phase of the hopping amplitude between the wire and Shiba bands becomes inhomogeneous.  
 One can invoke other ways of breaking the effective TRS. 
In presence of such TRS breaking terms, one can always decompose the  Hamiltonian describing the low-energy physics of this system of $H_{\rm sys}^{\rm eff}=H^{\rm eff}+H^{\rm eff}_{\rm TRSB}$ where $H^{\rm eff}_{\rm TRSB}$ contains all terms that break the effective TRS {\it i.e.} that make $H^{\rm eff}$ complex. $H_{\rm sys}^{\rm eff}$ then belongs to class D which is the initial class of the system under consideration.  Class D is characterized by  a $\mathbb{Z}_2$ invariant which means the 
system has one or zero  MBS at its extremity. Therefore the MBS in the ferromagnetic wire and in the Shiba band can hybridize in which case the topological character is lost.
This is reminiscent of  multibands  nanowires where an even number of occupied subbands realizes a trivial state while an odd number of occupied subbands  realizes a non-trivial topological state with MBS localized at its ends.\cite{Lutchyn2011,Stanescu2011}

\subsection{Dispersion relations of the low-energy Hamiltonian}

In order to characterize the topological properties of the Hamiltonian in Eq. (\ref{eq:heff}) and thus the number of MBS at one extremity of the chain, we can simply diagonalize the tight binding Hamiltonian and count the number of Majorana edge states. However, this simple strategy does not always allow to distinguish a MBS from another bound state occuring near the middle of the gap. We are to consider other bulk characterization of the topological properties. Therefore, we study Eq. (\ref{eq:heff}) with periodic boundary conditions such that momentum is conserved along the chain.
We introduce the Nambu spinor as:
\begin{equation}
\phi^T(k)=  (a_k , c_k, a^\dagger_{-k}, c^\dagger_{-k})~,
\end{equation}
with $k\equiv k_x$ the momentum along the chain, $a_k$ destroys an electron in the polarized conduction wire conduction band and $c_k$ destroys an electron in the polarized Shiba band. In this basis, $H^{\rm eff}= \sum_{k>0} \phi^\dagger(k) \mathcal{H}^{\rm eff}(k) \phi(k)$
with
\begin{equation} \label{eq:hk}
\mathcal{H}^{\rm eff}(k)= \hat h(k) \otimes\tau_z + \hat\Delta(k)\otimes\tau_x ,
\end{equation}
where the Pauli matrices $\tau$ denote the particle-hole space and 
\beq
\hat h(k)=\begin{pmatrix}
h^{a}(k)& M(k)\\
M(k)& h^c(k)
\end{pmatrix},~
\hat \Delta(k)=
\begin{pmatrix}
\Delta^a(k)& N(k)  \\
N(k)& \Delta^c(k)  
\end{pmatrix}.
\end{equation}

We introduced
\begin{eqnarray}
M(k)&=& -t\sqrt{\frac{\Delta}{2JS} } \left(   1+ \frac{2}{\sqrt{\pi k_Fa}}\sum_{m>0} \frac{1}{\sqrt{m}} \cos(k_Fma)\right.\nn\\
&\times&
\left.\cos(k_hma)\cos(kma)   e^{-ma/\xi} \right)=M(-k), \\
N(k)&=&t \sqrt{\frac{2\Delta}{JS} } \frac{1}{\sqrt{\pi k_Fa}}\sum_{m>0} \frac{1}{\sqrt{m}} \sin(k_Fma)\nn\\
&\times&\sin(k_hma)\sin(kma)   e^{-ma/\xi} =-N(-k) ,
\end{eqnarray}
and
\begin{eqnarray}
	h^{(c)}(k) &=& \epsilon_S-\Delta\sqrt{\frac{2}{\pi k_Fa} } \sum_{m> 0} \biggr(\frac{1}{\sqrt{m}} \cos(k_Fma-\frac{\pi}{4}) \nn \label{eq:hc}
	\\ && \times 2\cos(k_hma) \cos (kma)e^{-\frac{ma}{\xi}}\biggl),\\
	\Delta^{(c)}(k)&=&
	+\Delta\sqrt{\frac{2}{\pi k_Fa} } \sum_{m>0}\biggr(\frac{2}{\sqrt{m}} \sin(k_Fma-\frac{\pi}{4}) \nn\\
&&\times  \sin({kma})\sin(k_hma)e^{-\frac{ma}{\xi}} \biggl),\\
	h^{(a)}(k)&=&(\mu_w-J'S )-t_w\cos(k_ha)\cos(ka),\\
	\Delta^{(a)}(k)&=&\biggr(\frac{1}{J'S}+\frac{1}{\mu_w}\biggl)\frac{t_wt^2\pi\nu_0\sin(ka)\sin(k_ha)}{2}	,
\end{eqnarray}
 denote the Fourier transform of the low energy hopping and pairing Hamiltonian for the $a$ and $c$-electrons. In Eq. \eqref{eq:hc}, we remind that, in the deep Shiba limit, $\epsilon_S\approx\Delta(1-\alpha)$.

Note that the fact we have been able to write the effective Hamiltonian in \eqref{eq:hk} that way, requires the assumptions $N(k)=-N(-k), ~M(k)=M(-k), ~h(k)=h(-k)$ which  holds only if the effective TRS is present.

\begin{figure}[h]
\centering
\includegraphics[width=0.95\linewidth]{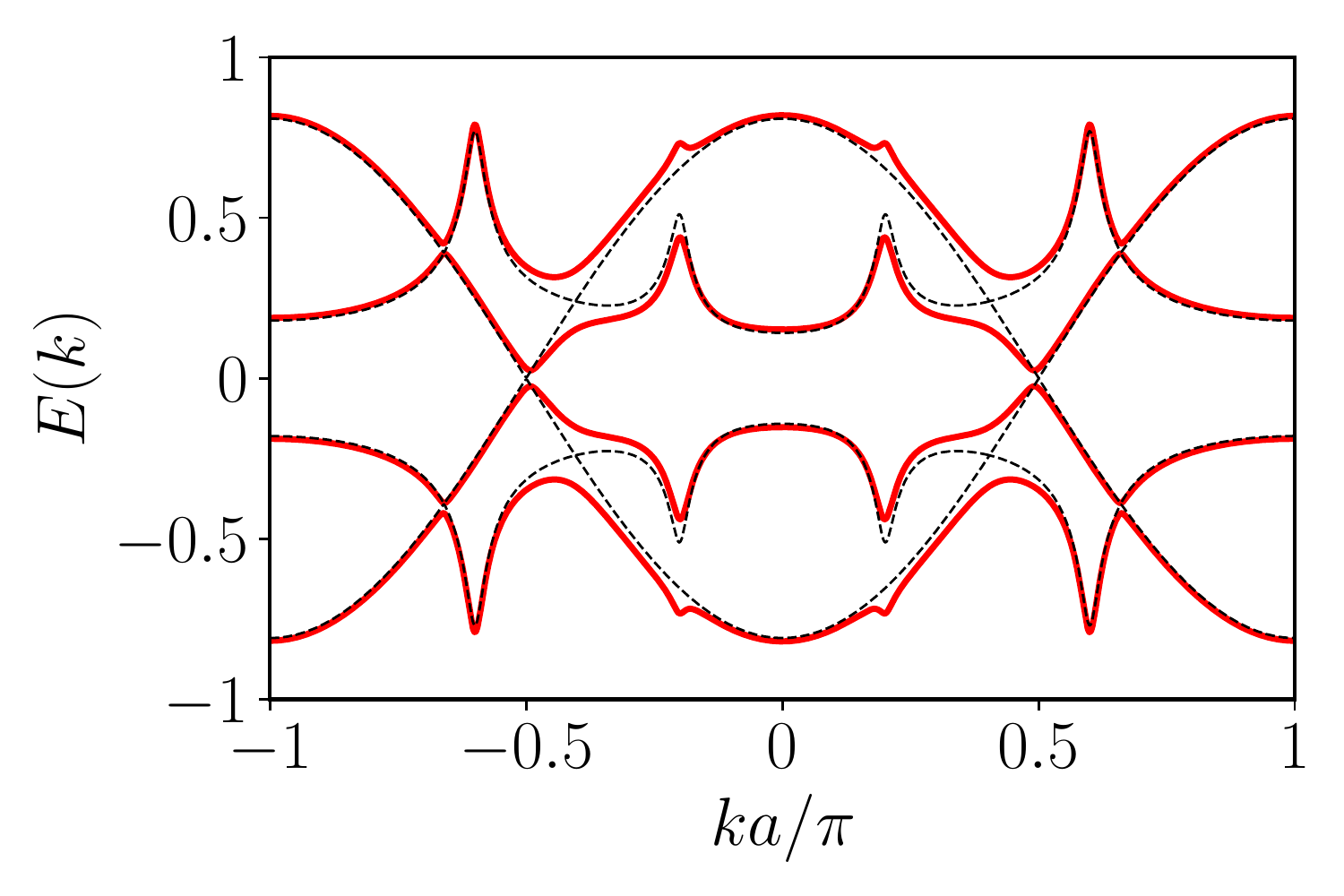}
	\caption{(color on line) A typical energy spectra (red solid line) associated with the low-energy Hamiltonian. The black dashed line shows a the uncoupled case$t=0$. The mid-gap bands correspond mainly to the Shiba band while the other two bands 
are built mainly from the wire. The peaks in the spectrum correspond to momenta $|k_F+k_h|$, $|k_F-k_h|$, $\mod(2\pi a)$. We can see the peaks also in the wire bands because of the hybridization. The used parameters are $\pi \nu_0=0.1$, $\Delta=1$, $k_ha/\pi=0.2$, $k_Fa/\pi=5.6$, $t_w=1$, $\epsilon_g=10$, $a/\xi=0.05$, $J'S=11$, $JS=11.5$ and $\alpha=1.15$, $t=0.5$.
}
	\label{fig:spectrum}
\end{figure}

The spectrum is obtained by diagonalizing numerically the Hamiltonian in Eq. (\ref{eq:hk}).  In what follows, we work with dimensionless units such that $\pi\nu_0=0.1$ and choose the range of parameters in the regime where the approximations that led to Eq. \eqref{eq:heff} are valid. An example of a typical spectrum is shown in Fig. \ref{fig:spectrum}.

\section{Topological properties and phase diagrams}
 In the previous section, we derived an effective Hamiltonian for a low-energy description of the system in the regime of interest $J\sim J'\gg \Delta$. In this section, we analyze its topological properties and  establish the phase diagram  of the system. 
Before, we introduce a few tools to characterize the topological properties.

\subsection{Winding number, parity and wave function}
{\em Winding number--} 
Let us first stress that   the winding number is defined only when the effective TRS is preserved.
Performing a first unitary transformation such that $a_k\longrightarrow e^{i\pi /4} a_k$, $c_k\longrightarrow e^{i\pi /4} c_k$,
$\mathcal{H}^{\rm eff}(k)\longrightarrow \hat h(k) \otimes\tau_z + \hat\Delta(k)\otimes\tau_y $
%
%
and then a rotation  $U=e^{-i\tau^y\pi/4}$, the new Hamiltonian reads:
\begin{equation}
\mathcal{H}'(k)= 
\begin{pmatrix}
0 & A(k)\\
A^\dagger(k) & 0
\end{pmatrix}.
\end{equation} 
In this basis, the Hamiltonian anti-commutes with $\tau^z$.
The winding number can be expressed as\cite{Tewari2012} 
\begin{equation}
\begin{aligned}
w=\int_0^\pi \frac{dk}{2\pi i}tr[\tau_z \mathcal{H} (k)\partial_k  \mathcal{H}^{-1}(k)].
\end{aligned}
\end{equation}
Introducing $z(k)=\frac{1}{\det{A(k)}}$, then
\begin{equation}
\label{eq:wphase}
\begin{aligned}
w=&\int_0^\pi \frac{dk}{\pi} \Im (\partial_k  \log (z(k))) 
= \frac{\arg(z(\pi))-\arg(z(0))}{\pi}  .
\end{aligned}
\end{equation}

{\em Parity--} Another important criteria to analyze the topological properties of the Hamiltonian is the parity operator $\mathcal{P}$. We remember that the relation between the parity and the winding number simply reads $\mathcal{P}=(-1)^w$. Because the pairing terms do not change the parity, $\mathcal{H}^{\rm eff}(k)$ in Eq. \eqref{eq:hk} and $\hat h(k)$ share  the same parity.
 Diagonalizing the electronic part of $\hat h(k)$,  we get two bands $E_+(k), E_-(k)$ characterized by a parity index $\mathcal{P}^\pm$. In that case,  
\beq
\mathcal{P}^\pm={sign(E_\pm(0)E_\pm(\pi))}.
\eeq
 Thus, if one band supports a single MBS, that band must cross the Fermi level $\mu$ an odd number of times. Therefore,  if $\mathcal{P}^++ \mathcal{P}^-=-1$, then the system is in a non-trivial topological phase.
However, if $\mathcal{P}=+$, this shows that we have an even number of MBS, {\it i.e.} $w=0,2,4\dots$.

{\em Nambu wave function--} Finally, in order to analyze the MBS, as mentioned above, we can simply diagonalize the tight binding Hamiltonian in Eq. \eqref{eq:heff} with open boundary conditions. For a lattice site labeled by $n$, we can define the 
Nambu wave function as :
$	\Psi_n=\left(u_n^{(a)}, v_n^{(a)}, u_n^{(c)}, v_n^{(c)}\right)$ such that the
 probability to find an electron (resp. hole) in the band $S=c,a$ is simply 
$|	u_n^{(S)}|^2$ (resp. $|	v_n^{(S)}|^2$).

\begin{figure}[t]
	\centering
	\begin{tabular}{cc}
		\includegraphics*[width=0.48\columnwidth]{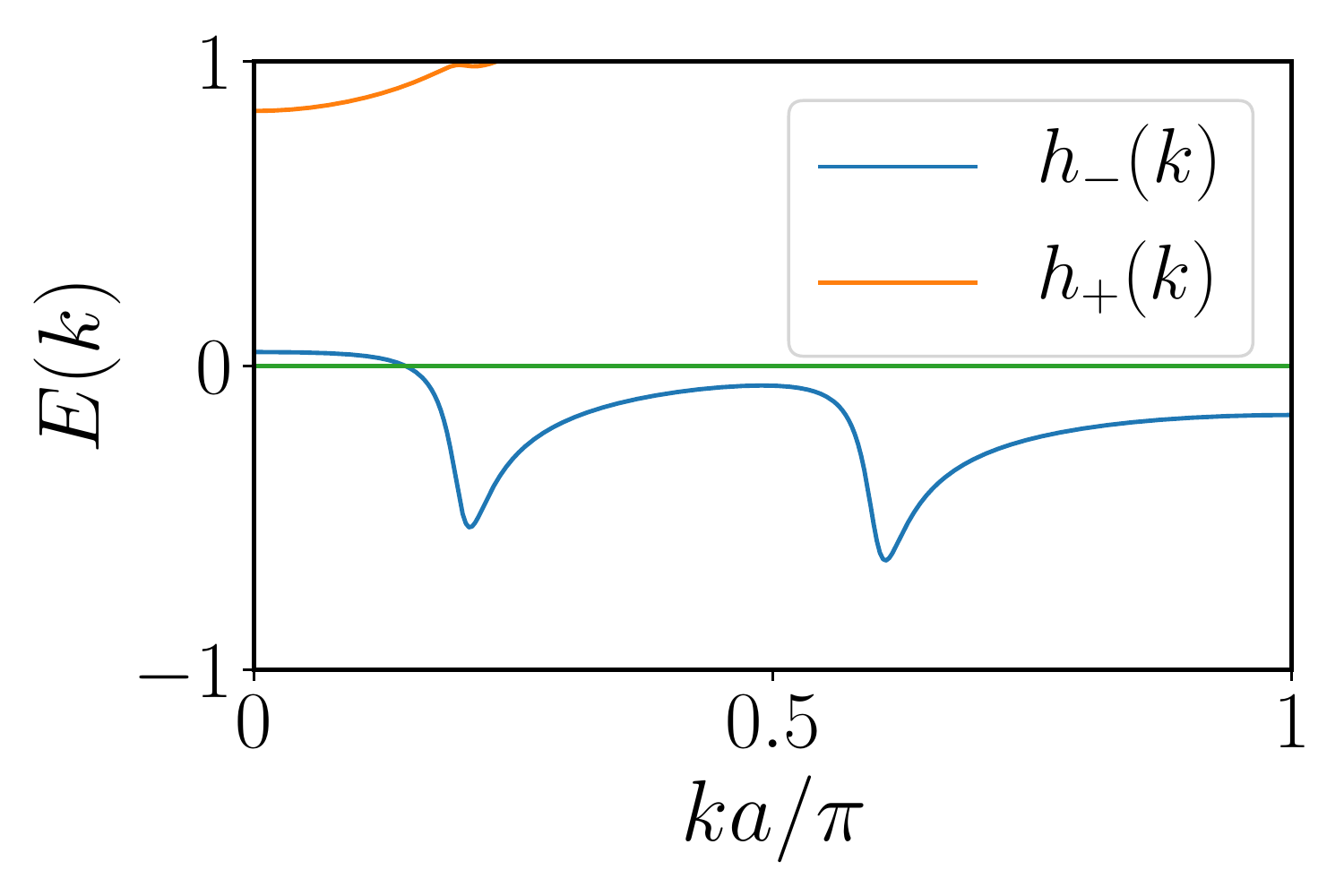} & 
		\includegraphics*[width=0.48\columnwidth]{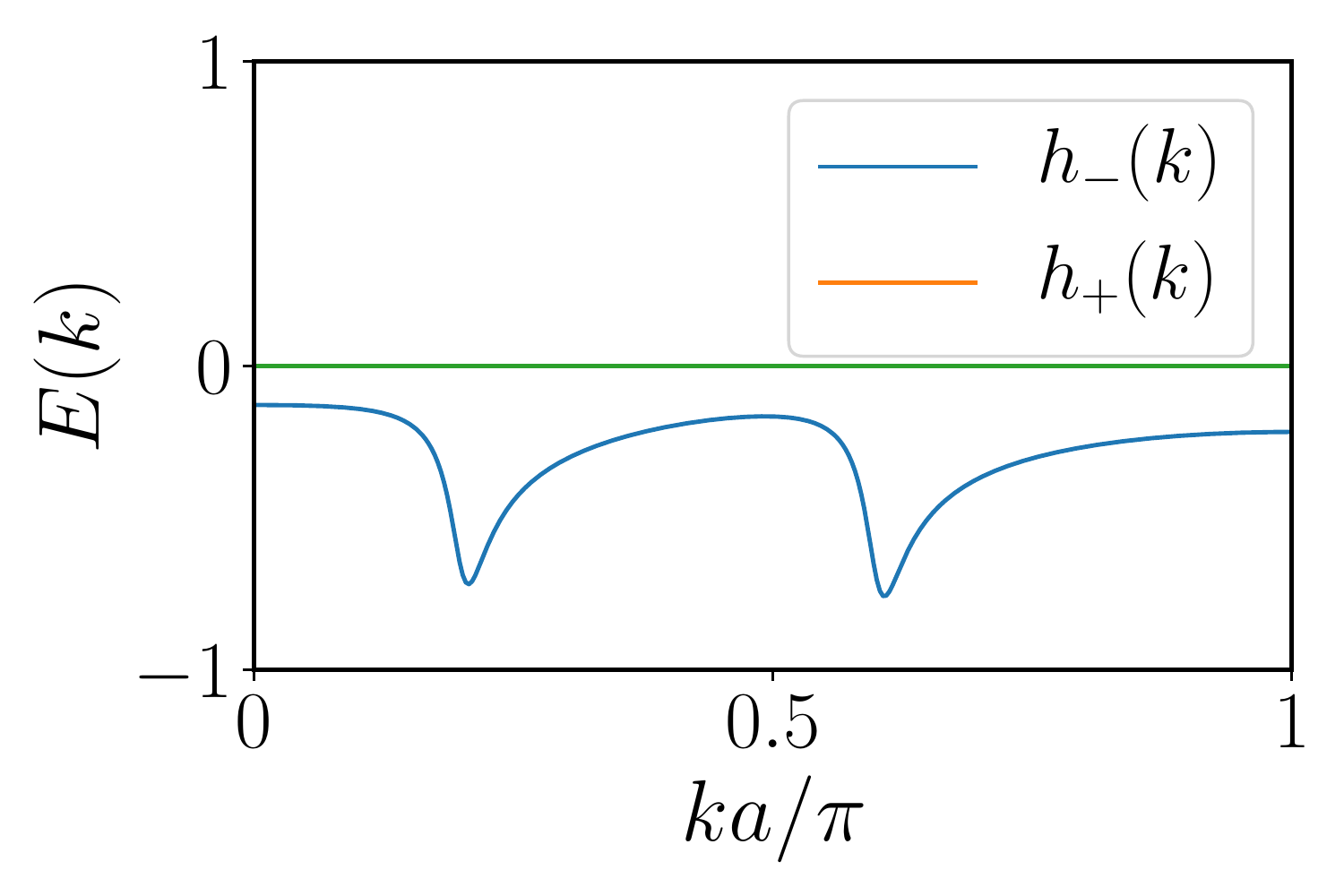}  \\
		\includegraphics*[width=0.48\columnwidth]{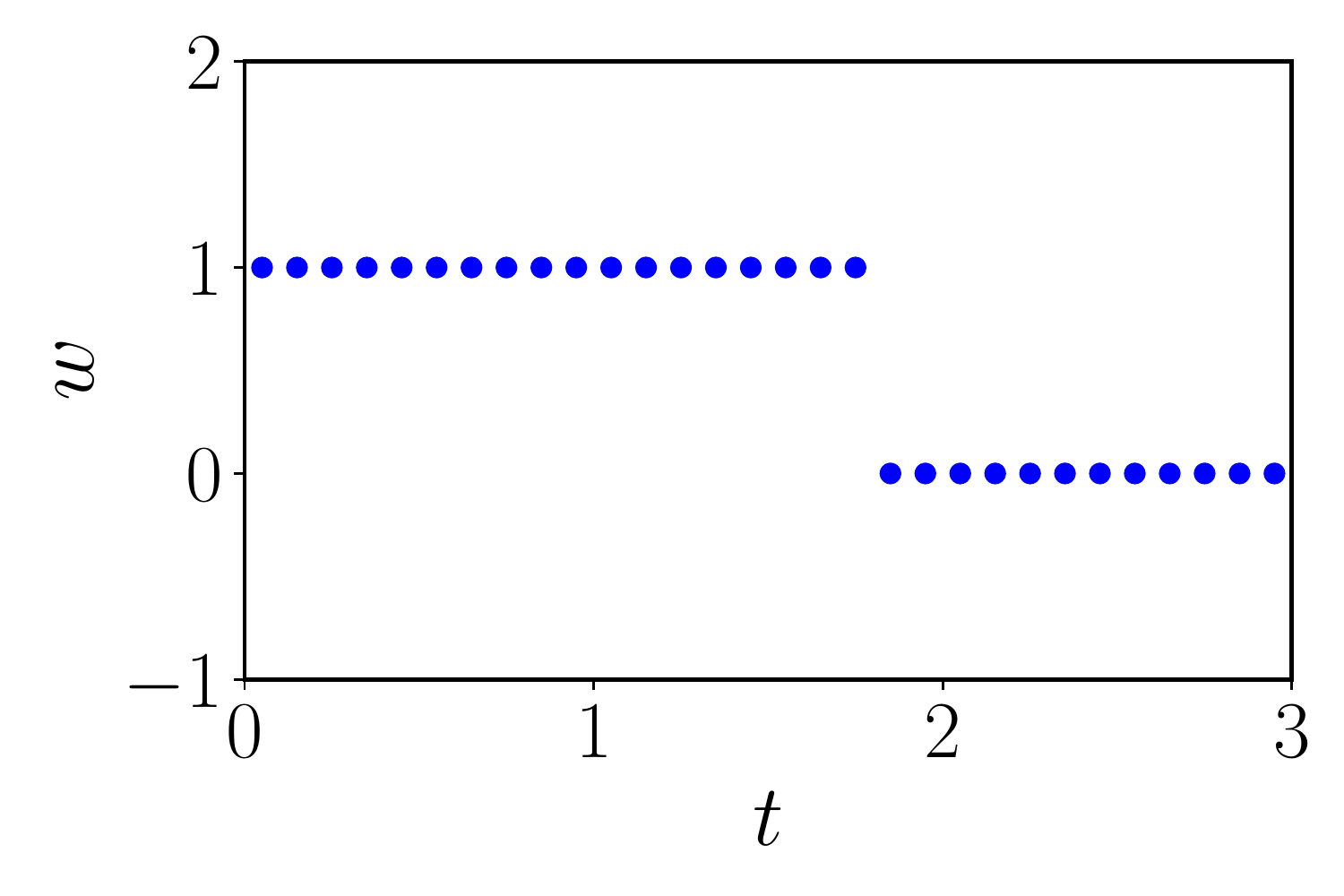} & 
		\includegraphics*[width=0.49\columnwidth]{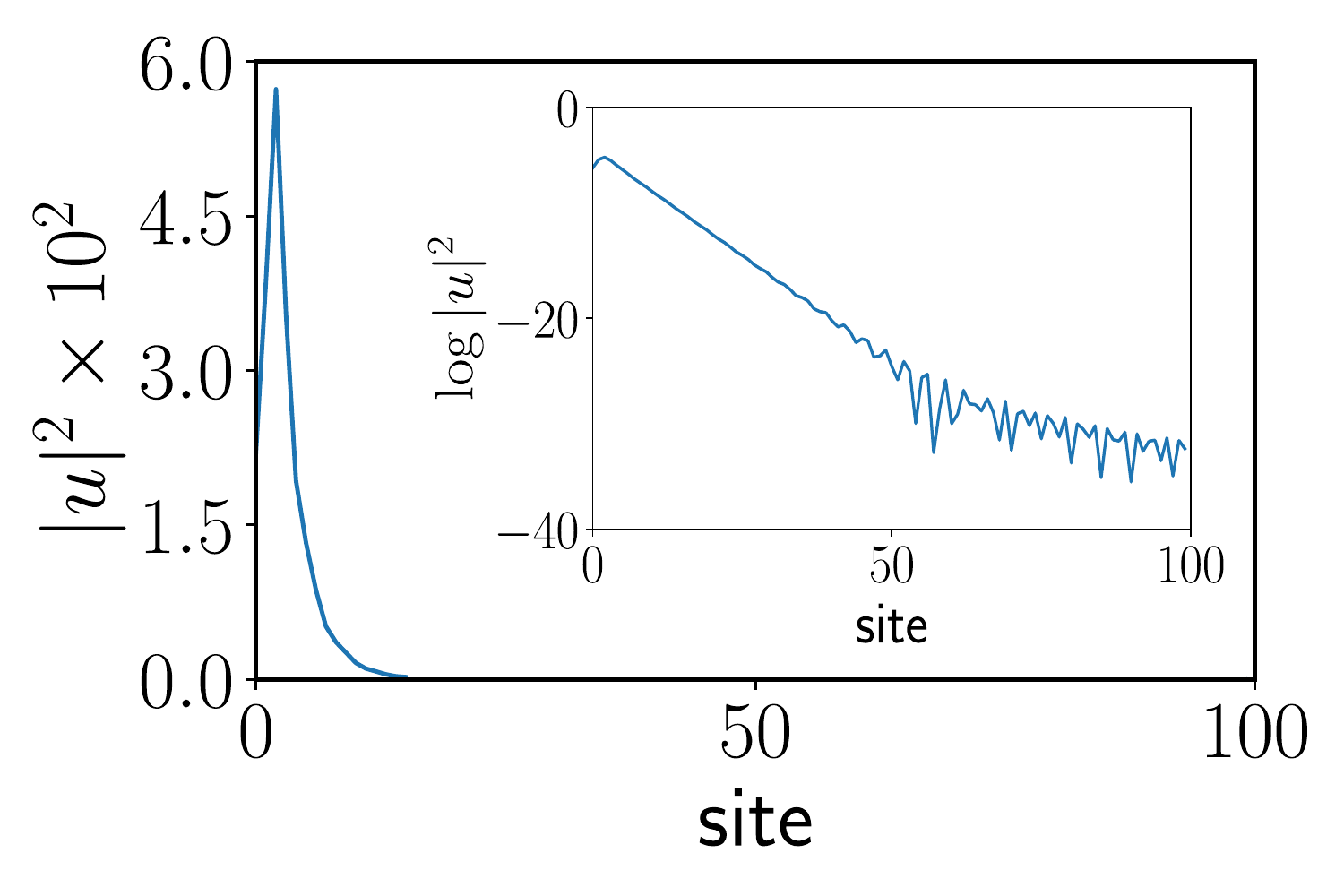}  
\end{tabular}
	\caption{(Color on line) 
		Upper row: Plot of the two bands of the normal part of $\hat h(k)$  with $t=1.5$ (left) and $t=2.5$ (right).
We can see that there  is one  crossing at the Fermi level for the left panel  and no crossing in the right panel. Therefore, we expect  the Shiba (hybridized) band to be topological with $|w|=1$.
Lower row: In the left panel, we plot the winding Number as a function the hybridization parameter $t$ confirming that for $t=1.5$, the Shiba  (hybridized) band has a non-trivial topology. For open periodic conditions, the Shiba band has a localized MBS. We plot on the right panel the  electronic part of its wave function $|u|^2$ in the Shiba band for $t=1.5$ and $N=200$ sites. The inset represents a log plot of the same quantity suggesting  two different exponential decay lengths. Similar extent of the wave function is found in the wire due to the hybridization between the two bands.
The used parameters are 
		$\pi \nu_0=0.1$, $\Delta=1$, $k_ha/\pi=0.2$, $k_Fa/\pi=5.6$, $t_w=1$, $\epsilon_g=10$, $a/\xi=0.05$, $J'S=9.5$, $JS=11$ and $\alpha=1.1$.
}
\label{fig-Shiba}
\end{figure}

\subsection{Topological properties}
As stressed before, we are interested in the regime where $J\sim J'\gg \Delta$ and in the dilute regime $k_Fa\gg 1$. Although we reduced the {\it a priori} complicated system to the low-energy Hamiltonian in Eq. \eqref{eq:heff} (or Eq. \eqref{eq:hk} in k-space with periodic boundary conditions), $H^{\rm eff}$ still contains many parameters.
Before switching on the hybridization between the both bands,  three different situations can be encountered: i) only the Shiba band is topological  ii) only the wire conduction is topological and iii)  both bands can be topological.

\begin{figure}[h]
	\centering
	\begin{tabular}{cc}
		\includegraphics*[width=0.48\columnwidth]{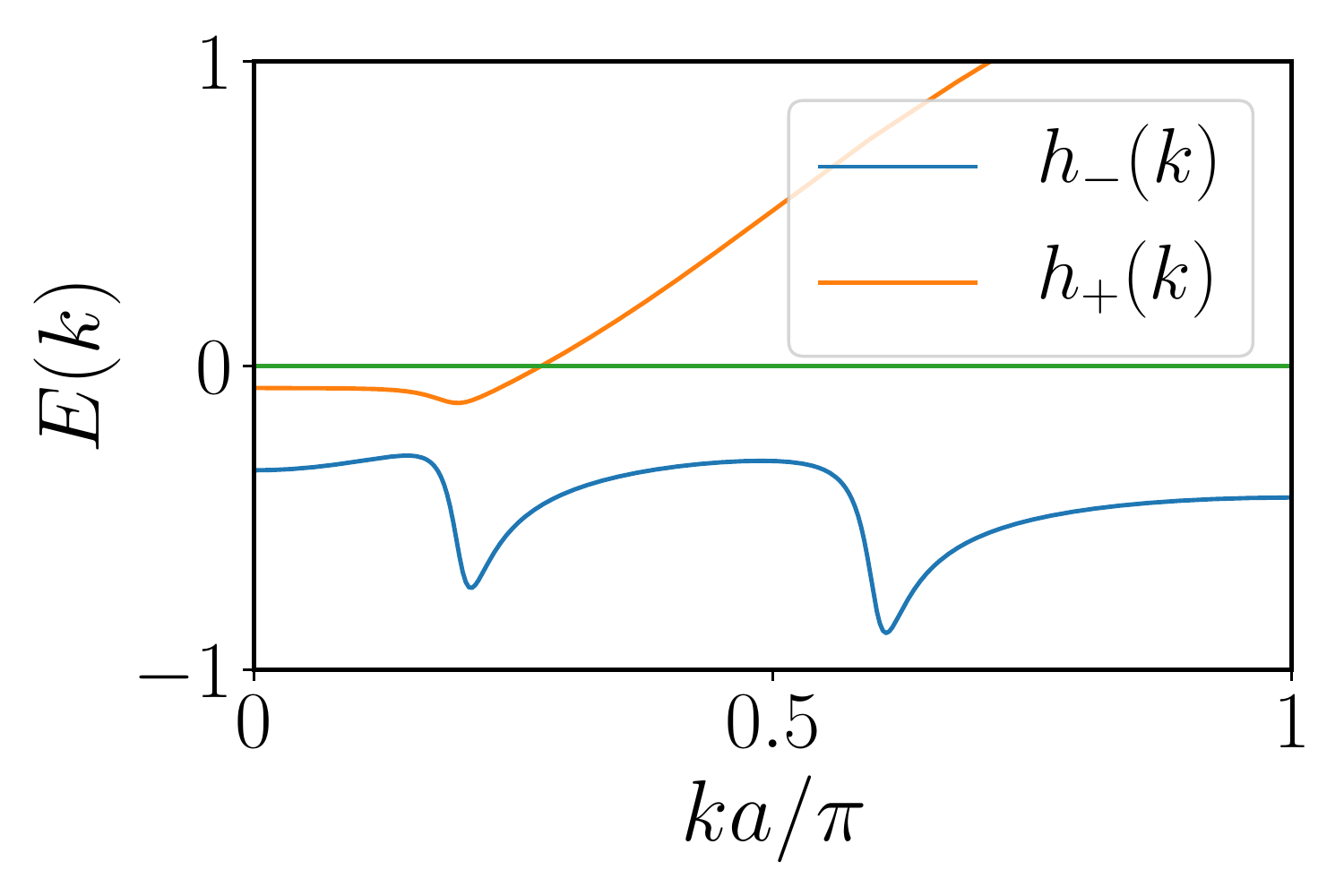} & 
		\includegraphics*[width=0.48\columnwidth]{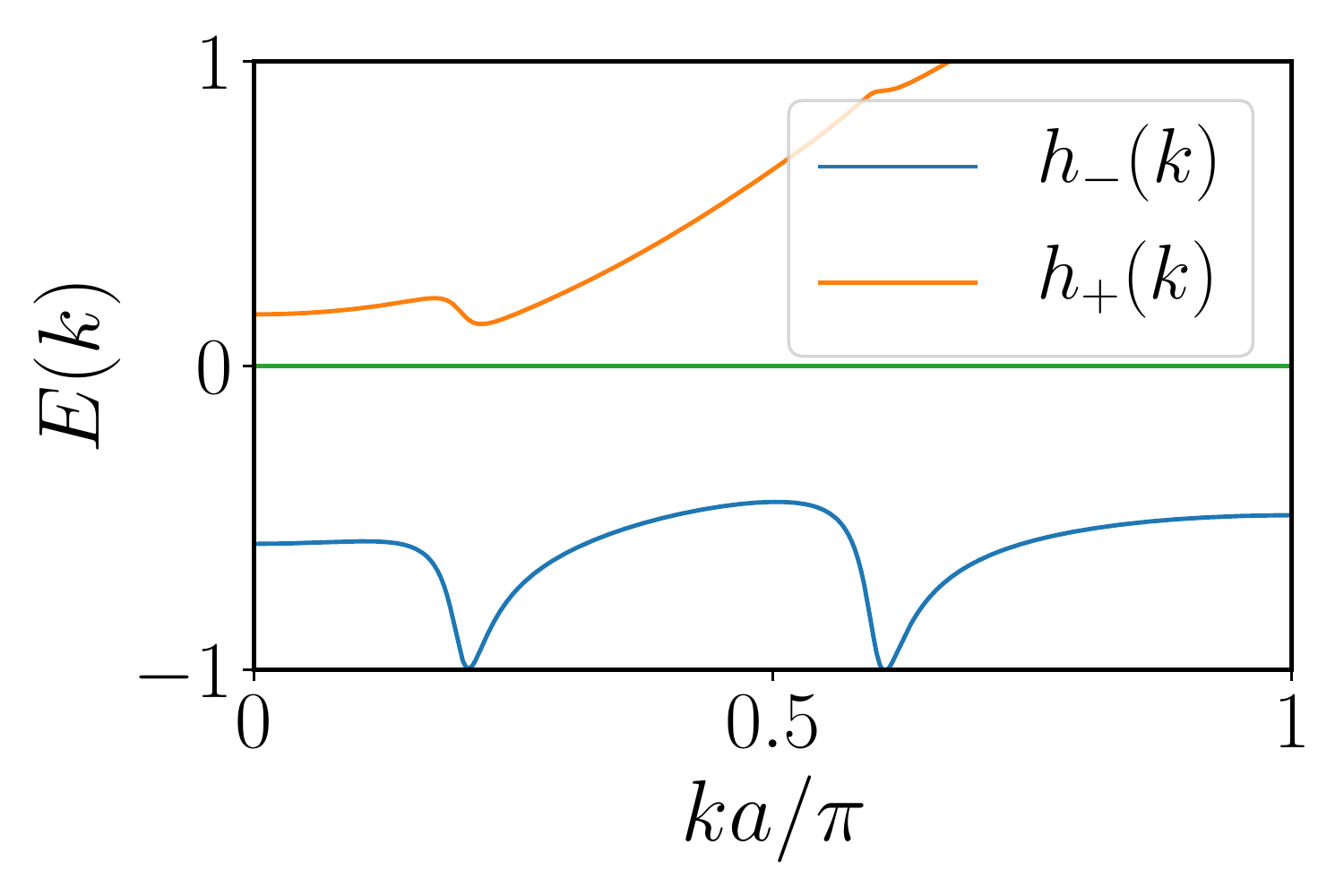}  \\
		\includegraphics*[width=0.48\columnwidth]{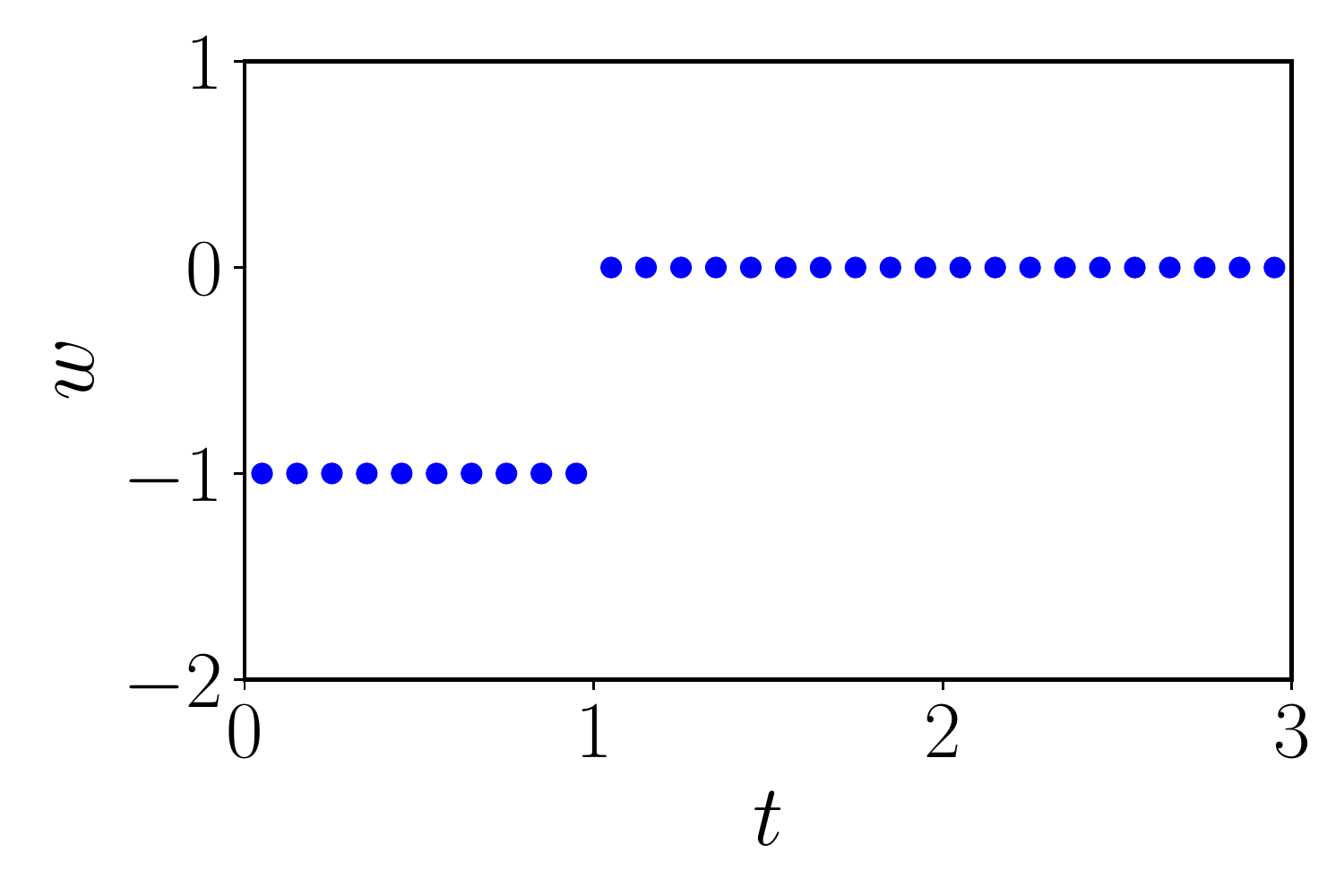} & 
		\includegraphics*[width=0.49\columnwidth]{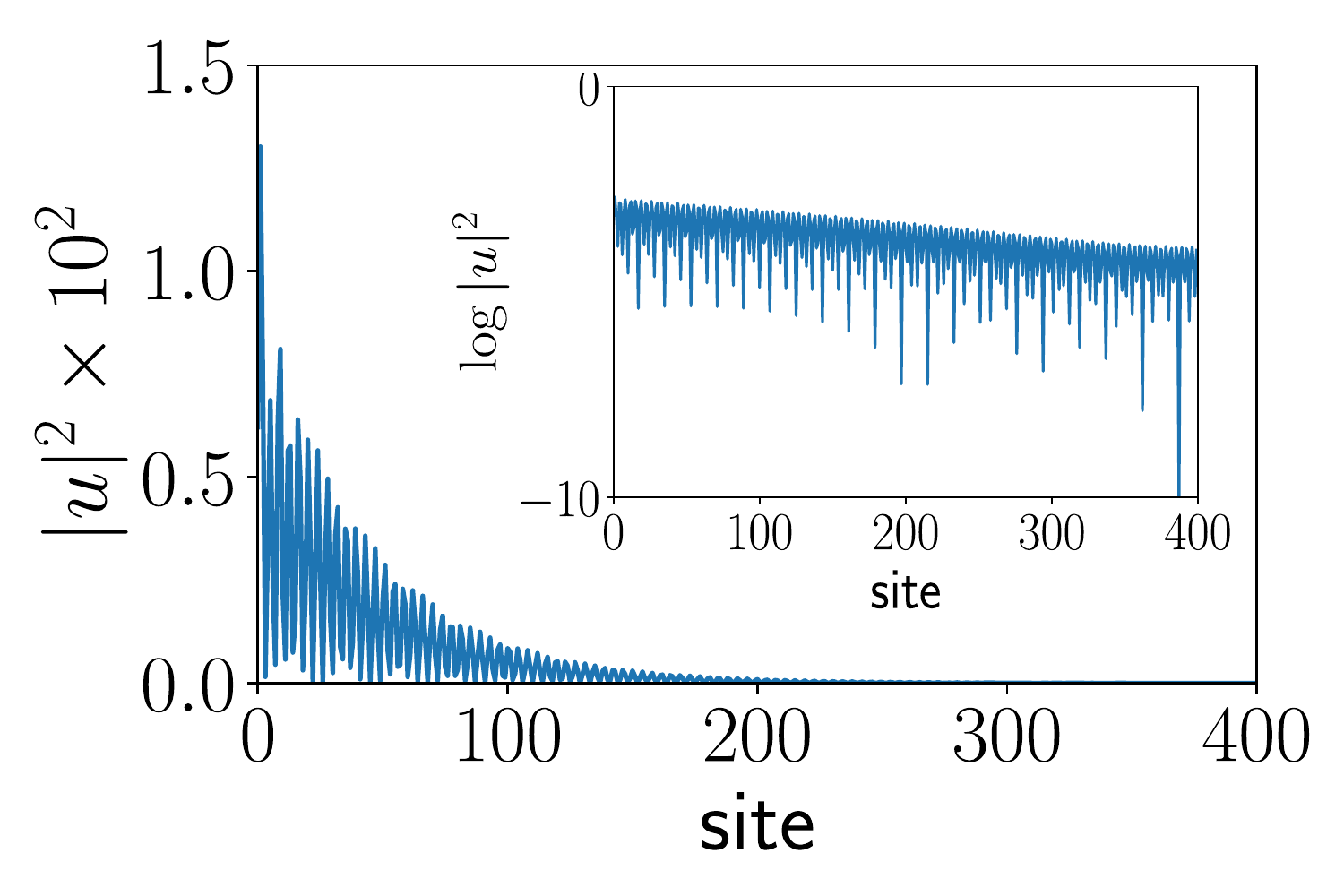}  
\end{tabular}
	\caption{(Color on line) 
		Upper row: Plot of the two bands of the normal part of $\hat h(k)$ with $t=0.5$ (left) and $t=2$ (right).
We can see that there  is one crossing at the Fermi level in the left panel  and no crossing on the right panel. Therefore, we expect  the wire (hybridized) band to be topological with a winding number $|w|=1$.
In the lower left panel, we plot the winding number as a function the hybridization parameter $t$ confirming that for $t=0.5$, the wire (hybridized) band has a non-trivial topology. For open periodic conditions, we plot on the lower right panel the  electronic part of its wave function $|u|^2$ in the wire band for $t=0.5$ and $N=800$ sites. The inset represents a log-plot of the same quantity showing the exponential decay. Similar spatial extent of the wave function is found in the Shiba band due to the small hybridization. The used parameters are $\pi \nu_0=0.1$, $\Delta=1$, $k_ha/\pi=0.2$, $k_Fa/\pi=5.6$, $t_w=1$, $\epsilon_g=10$, $a/\xi=0.05$ and $J'S=10.5$ , $JS=14$, $\alpha=1.4$.
	}
	\label{fig-wire}
\end{figure}

We start with  $JS=11$ and $J'S=9.5$ (the other parameters being indicated in the caption of the figures) which correspond in the absence of hybridization to the Shiba band supporting one MBS and the wire band being topologically trivial. 
In the upper row of Fig. \ref{fig-Shiba}, we plot the two bands of the normal part of $\mathcal{H}^{\rm eff}(k)$ in Eq. \eqref{eq:hk} for the hybridization parameter $t=1.5$ (left) and $t=2.5$ (right). In the left panel, one band crosses the Fermi energy while on the left panel no crossing is obtained. This indicates that for $t=1.5$ (weak hybridization), the Shiba band remains topological. Increasing $t$, the two bands strongly hybridize and the system becomes topologically trivial. This picture is confirmed by directly plotting the winding number as a function of $t$. Indeed for $t\geq 2$, the system becomes topologically trivial and no MBS is expected to be found for open boundary conditions. However for $t\leq 2$, we expect two MBS localized at each extremity of the chain (see lower right panel of Fig. \ref{fig-Shiba}). Though the wave function of one MBS is mainly built from the Shiba band, there is also a small part localized in the wire band.

\begin{figure}[h]
	\centering
	\begin{tabular}{cc}
		\includegraphics*[width=0.48\columnwidth]{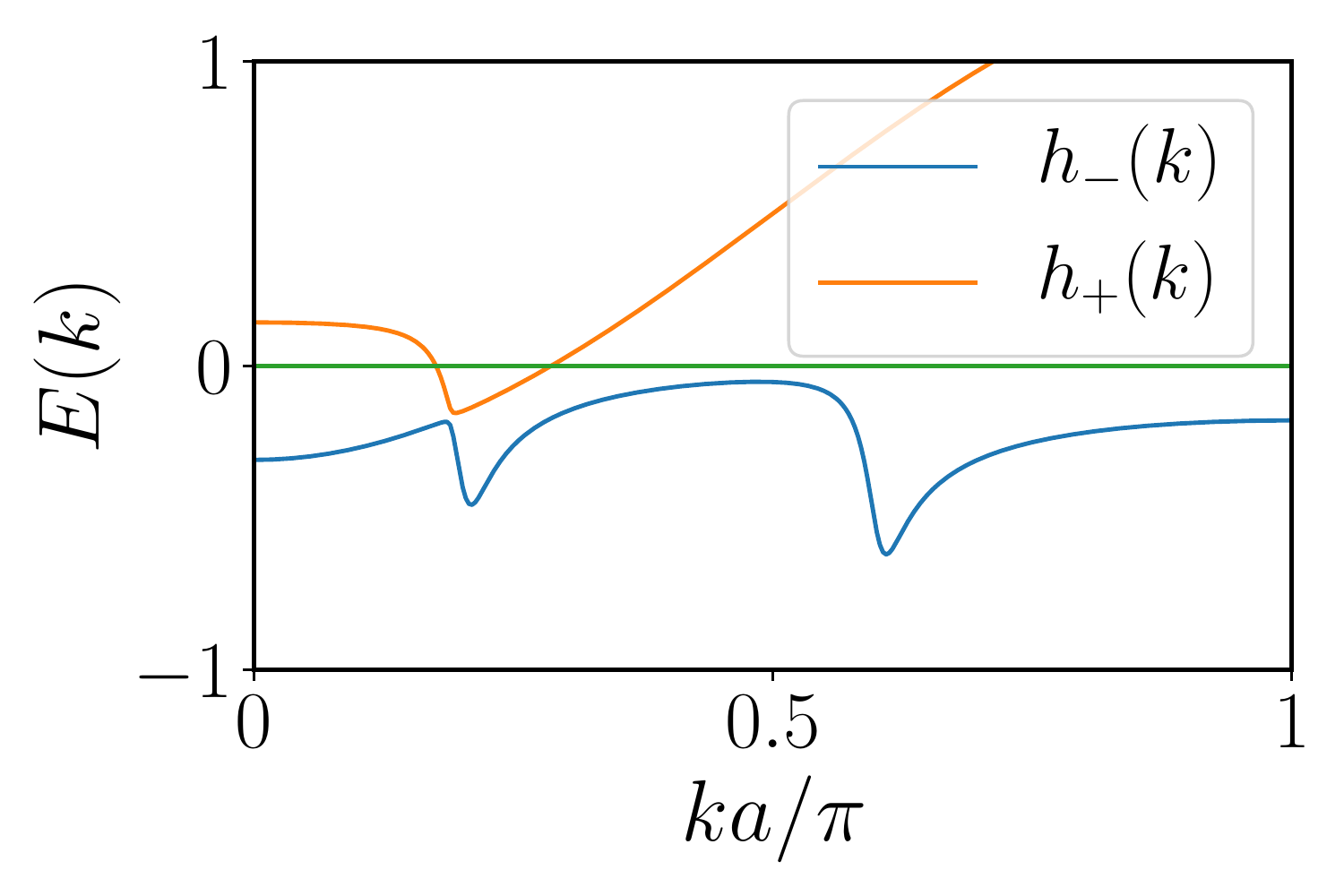} & 
		\includegraphics*[width=0.48\columnwidth]{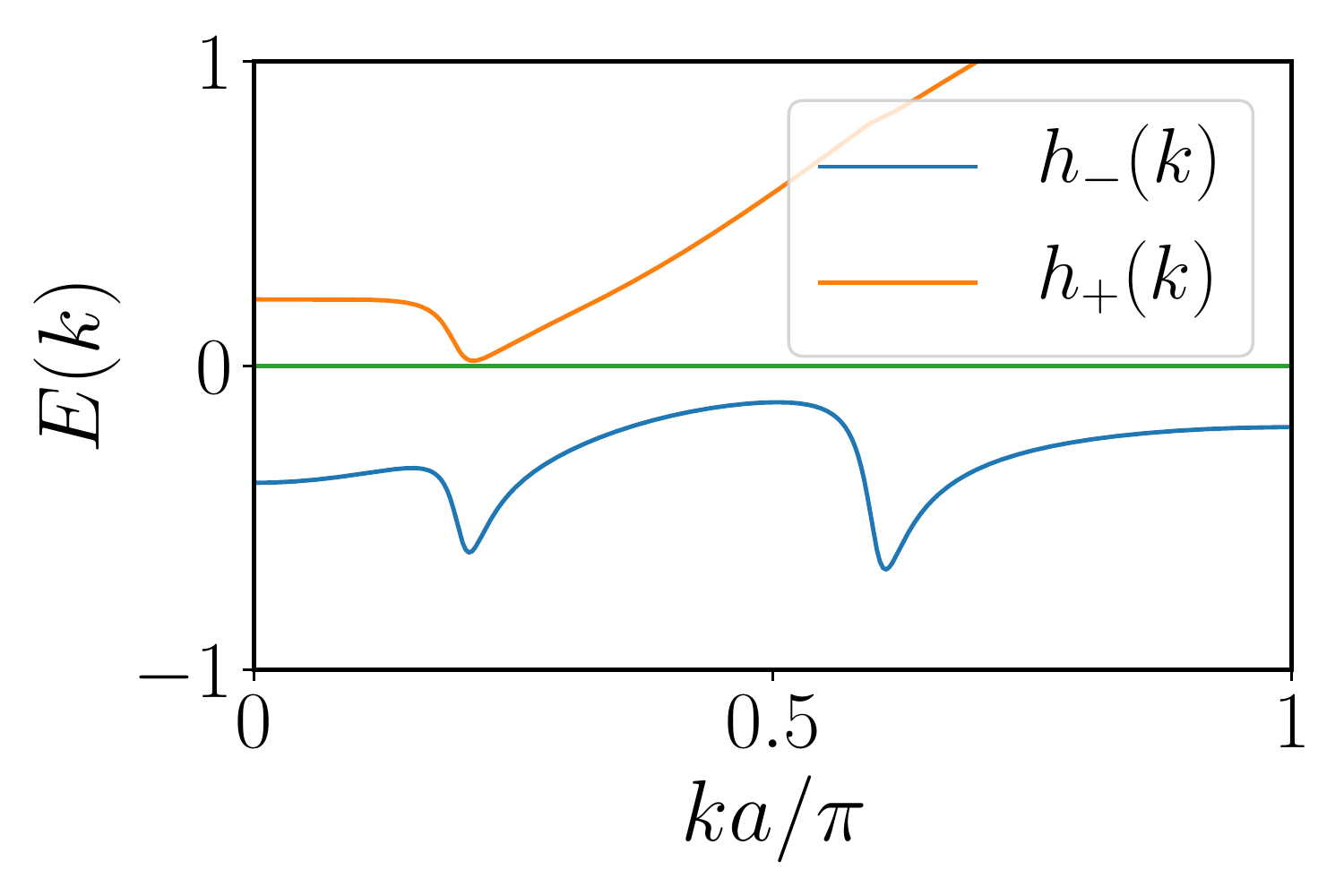}  \\
		\includegraphics*[width=0.48\columnwidth]{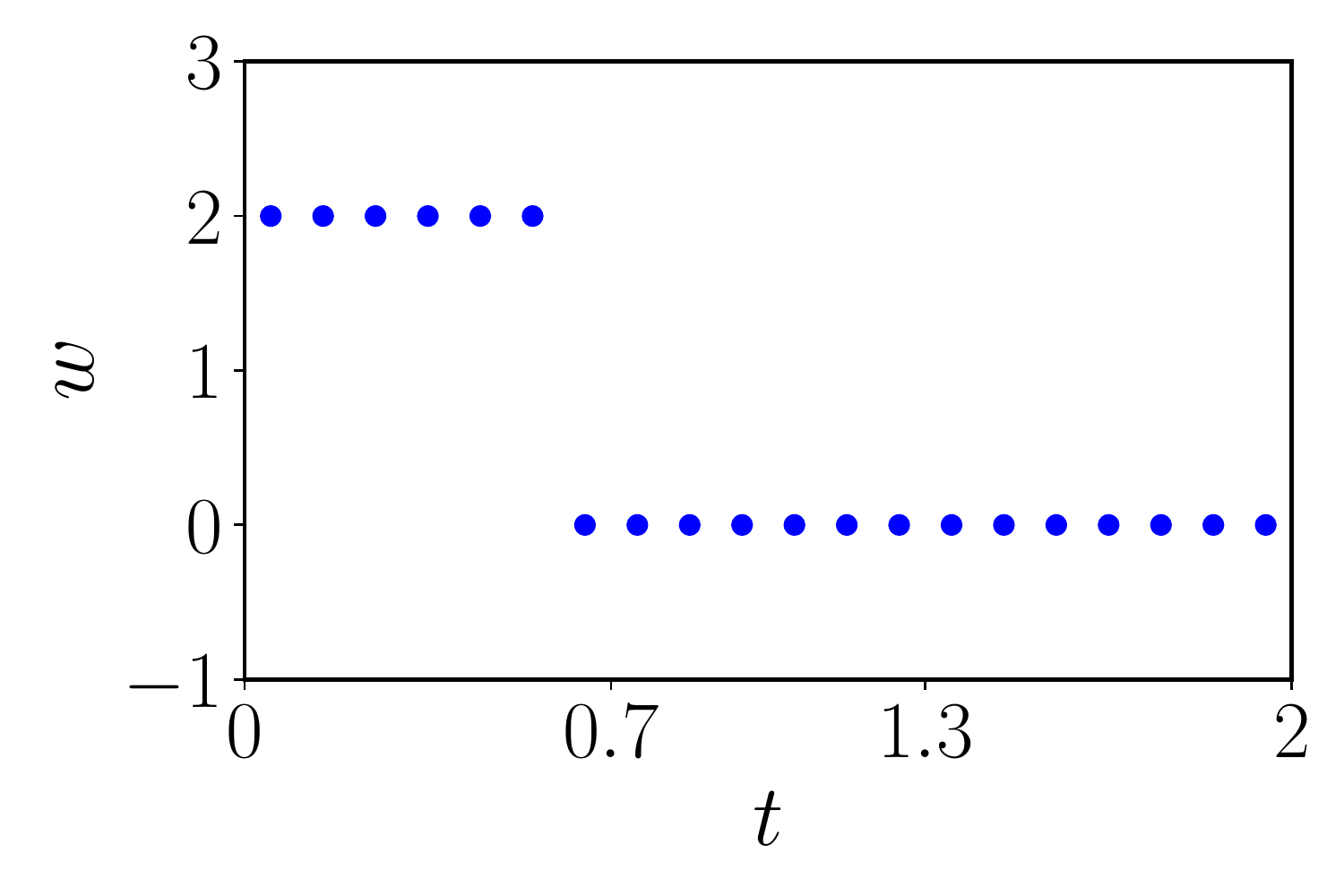} & 
		\includegraphics*[width=0.49\columnwidth]{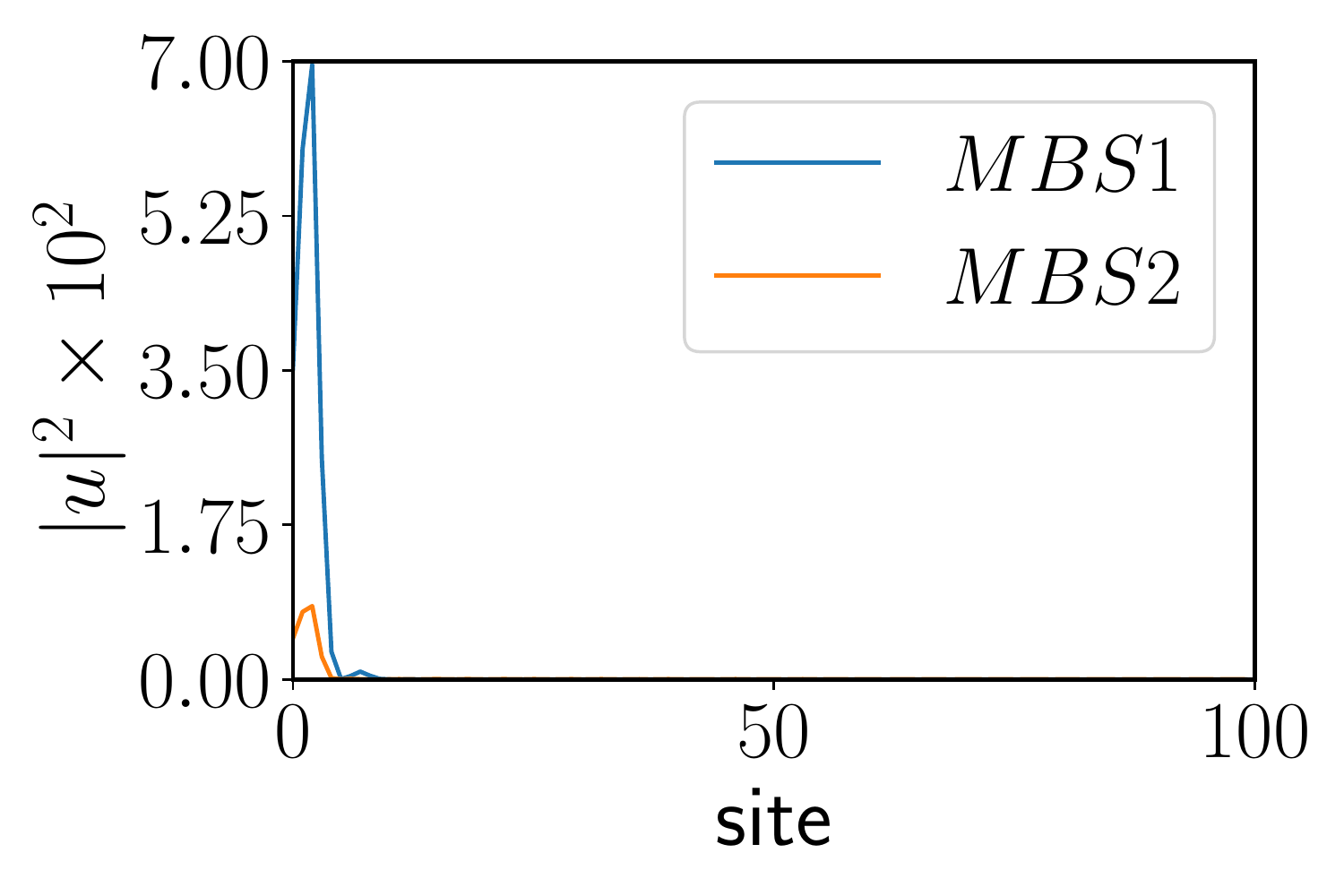}  
\end{tabular}
	\caption{(Color on line) 
		Upper row: Plot of the two bands of the normal part of $\hat h(k)$ with $t=0.1$ (left) and $t=1$ (right).
We can see that there  are two crossings at the Fermi level in the left panel  and no crossing on the right panel. 
In the lower left panel, we plot the winding Number as a function the hybridization parameter $t$. This implies that for $t=0.1$, the two bands have a non-trivial topology. For open periodic conditions, we plot on the lower right panel the  electronic part of the  wave function $|u|^2$ in the Shiba band for $t=0.1$ and $N=200$ sites. The used parameters are $\pi \nu_0=0.1$, $\Delta=1$, $k_ha/\pi=0.2$, $k_Fa/\pi=5.6$, $t_w=1$, $\epsilon_g=10$, $a/\xi=0.05$ and $J'S=10.5$ , $JS=11.5$.
}
	\label{fig-shiba-wire}
\end{figure}

For $JS=14$ and $J'S=10.5$, the wire band is topological at small $t$ while the Shiba band is normal. When switching on the hybridization, this situation holds up to $t\leq 1$ where there is topological transition. This is confirmed by plotting the spectrum of the bands for $t=0.5$ (upper left panel of Fig. \ref{fig-wire}) with one crossing and $t=2$ (upper right panel of Fig. \ref{fig-wire}) where no crossing is found. Indeed, for $t\geq 1$, the two bands become trivial and the total winding number is $w=0$ (lower left panel of Fig. \ref{fig-wire}). We plot the spatial extent of the MBS wave function in the wire band in the lower right panel of  Fig. \ref{fig-wire}  for $t=0.5$ and $N=800$ sites. A similar pattern is found in the Shiba band due to the hybridization. Note that the MBS wave function have a much larger spatial extent. This is due to the fact that the gap in the wire band is small for this set of parameters.
Furthermore, the gap is located at finite $k$ around $ k\approx \pm  \pi/4a $ which may explain the fast oscillations of the wavefunction.

For $JS=11.5$ and $J'S=10.5$, both bands are topological at weak hybridization. In the upper left panel of Fig. \ref{fig-shiba-wire}, we plot the spectrum for the normal part of the effective Hamiltonian for $t=0.1$ and found two crossings of the Fermi levels. Instead for $t=1$ (upper right panel of Fig. \ref{fig-shiba-wire}), no crossing is found. In both cases, the parity being even, no definite conclusion can be drawn at this level. To ascertain the non-trivial topology of the system, we directly plot the winding number in the lower left panel of  Fig. \ref{fig-shiba-wire} and find $w=2$ for $t\lesssim 0.65 $. We therefore expect one extremity of the chain to support two MBS.  The electronic part of the wire Majorana wave function is plotted in the lower right panel of Fig. \ref{fig-shiba-wire}. We clearly observe the two MBS  localized at one end of the chain, one MBS coming mainly from the wire band, the other MBS being shared with the Shiba  band due to hybridization.


\begin{figure}[h]
	\centering
	\begin{tabular}{cc}
		\includegraphics*[width=0.48\columnwidth]{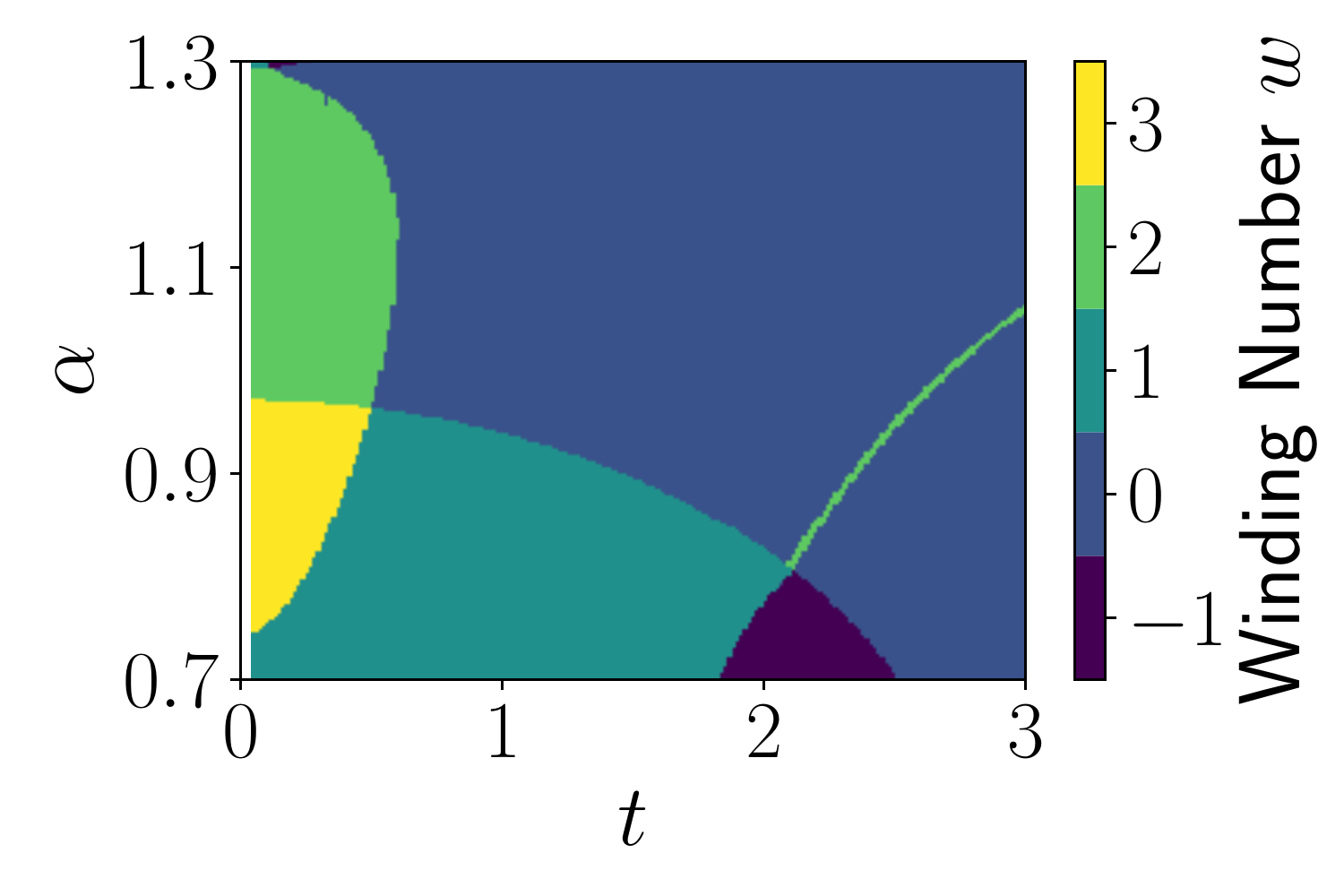} & \includegraphics*[width=0.48\columnwidth]{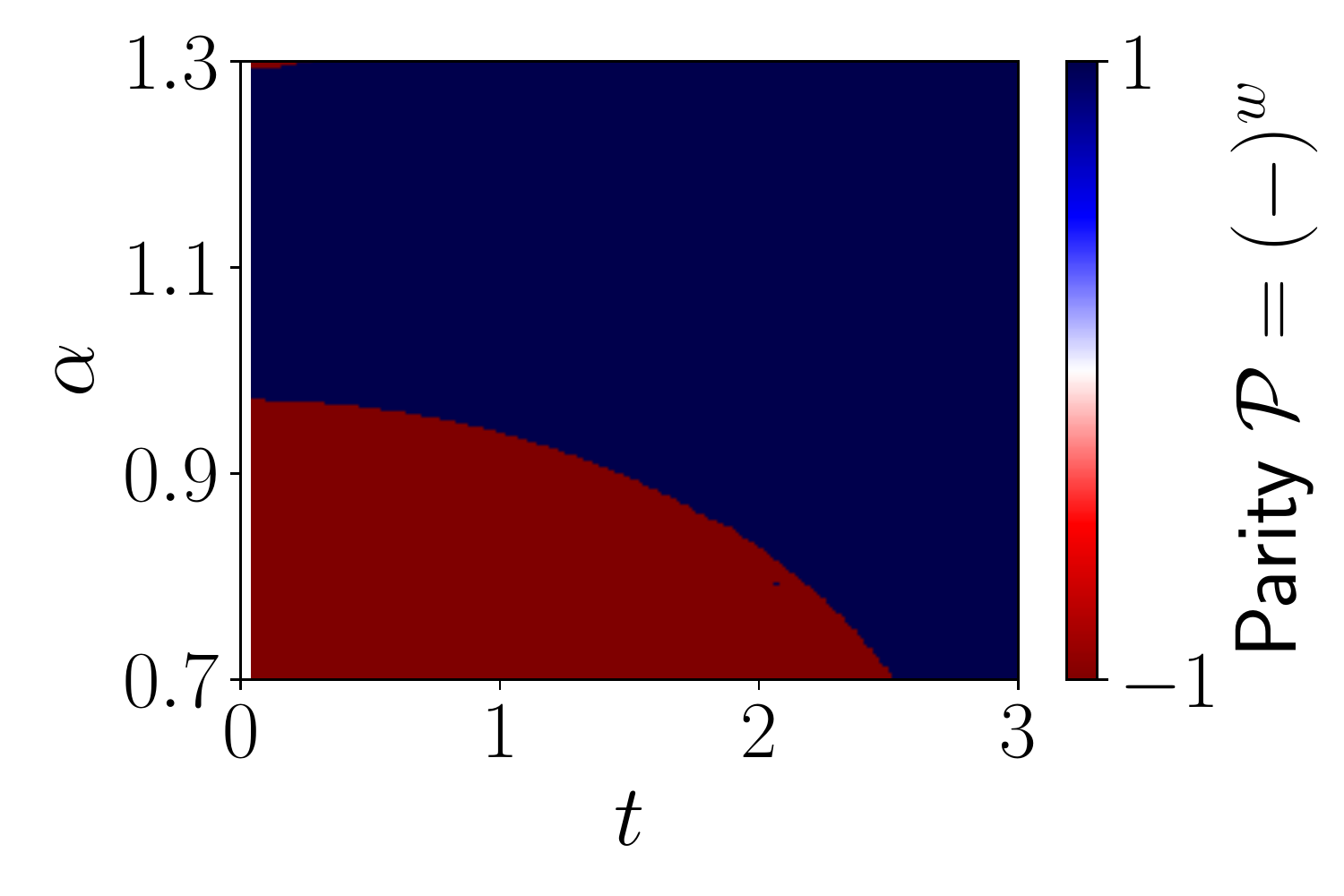}   \\
		\includegraphics*[width=0.48\columnwidth]{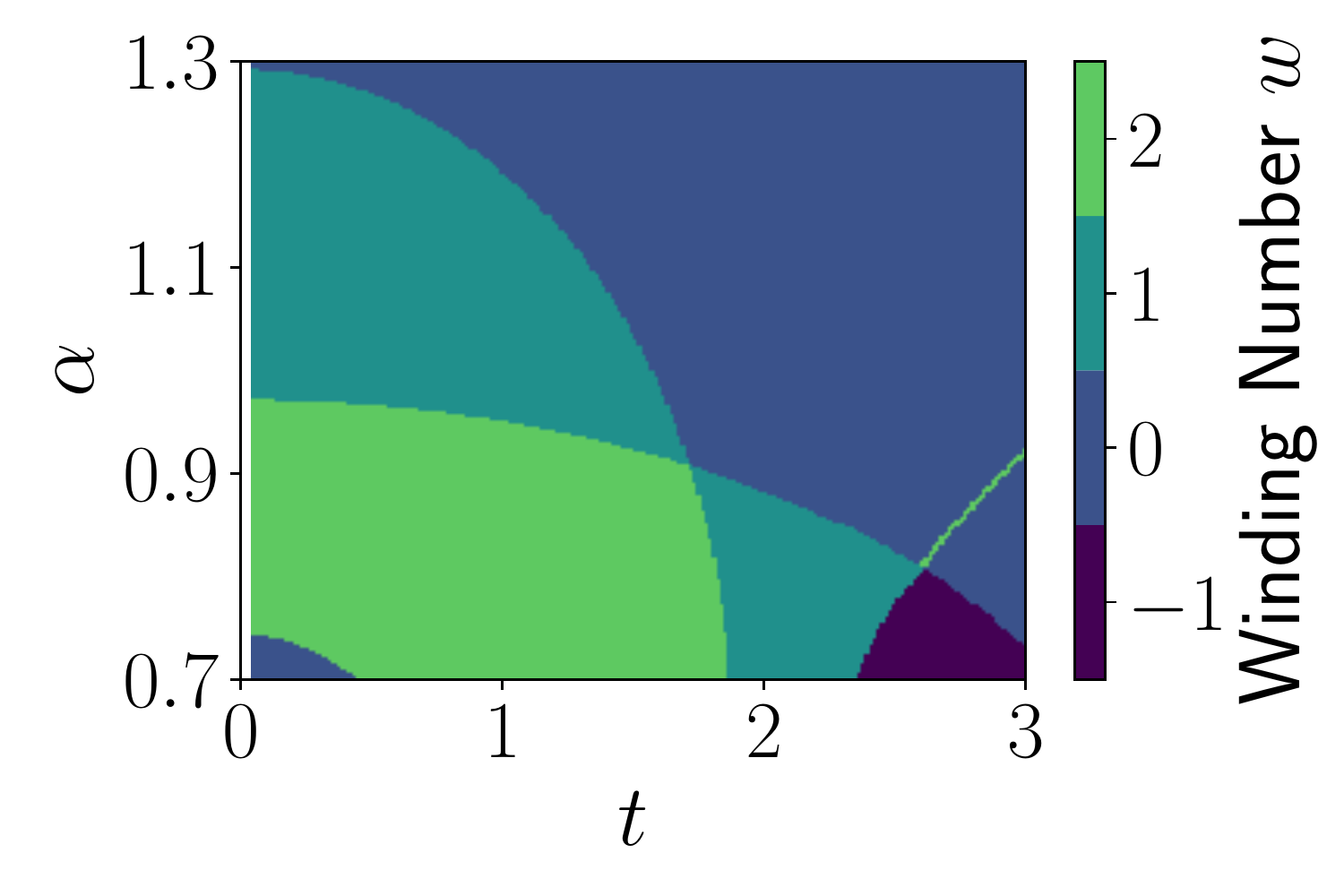} & \includegraphics*[width=0.48\columnwidth]{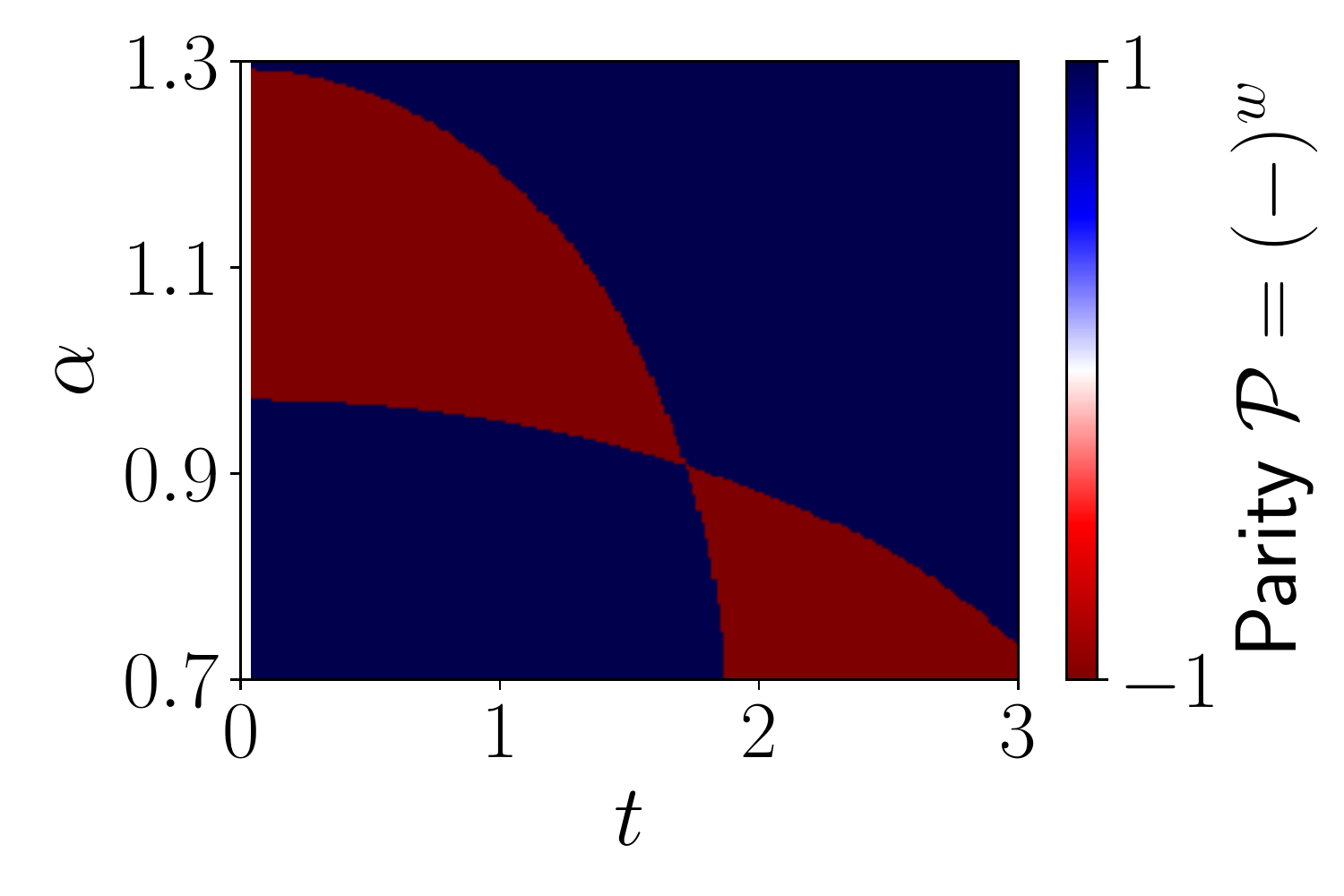}
\end{tabular}
	\caption{(Color on line) 
Winding number $w$ (left column) and parity $\mathcal{P}$ (right column) as a function of $t$ and $\alpha=\pi \nu_0 JS$. The color code indicate the possible integer values that the winding number $w$  and parity $\mathcal{P}$ can take. The used parameters are $\pi \nu_0=0.1$, $\Delta=1$, $k_ha/\pi=0.2$, $k_Fa/\pi=5.6$, $t_w=1$, $\epsilon_g=10$, $a/\xi=0.05$ and $J'S=10.5$ (upper row), $J'S=9.8$ (lower row).
		}
	\label{fig-phase-diagram1}
\end{figure}

\subsection{Phase diagrams}

In the previous section, we have exhibited  different cases where we can  have either $w=0,1,2$ by varying both $J$ and $J'$.
This can be summarized by plotting the winding number $w$ as a function of $t$ and $\alpha=\pi \nu_0 JS$ for fixed values of $J'S$.
This is shown in Fig. \ref{fig-phase-diagram1}. The winding number can  reach values up to $w=3$. Indeed the Shiba band can support up to two MBS \cite{Pientka2013} while the wire band can also have one MBS. Therefore at weak hybridization $t$, we can indeed reach phases with three MBS. The transitions between the different phases is characterized by a gap closing. At strong hybridyzation, the system becomes trivial when both $J$ and $J'$ are comparable.

\begin{figure}[h]
	\centering
	\begin{tabular}{cc}
		\includegraphics*[width=0.48\columnwidth]{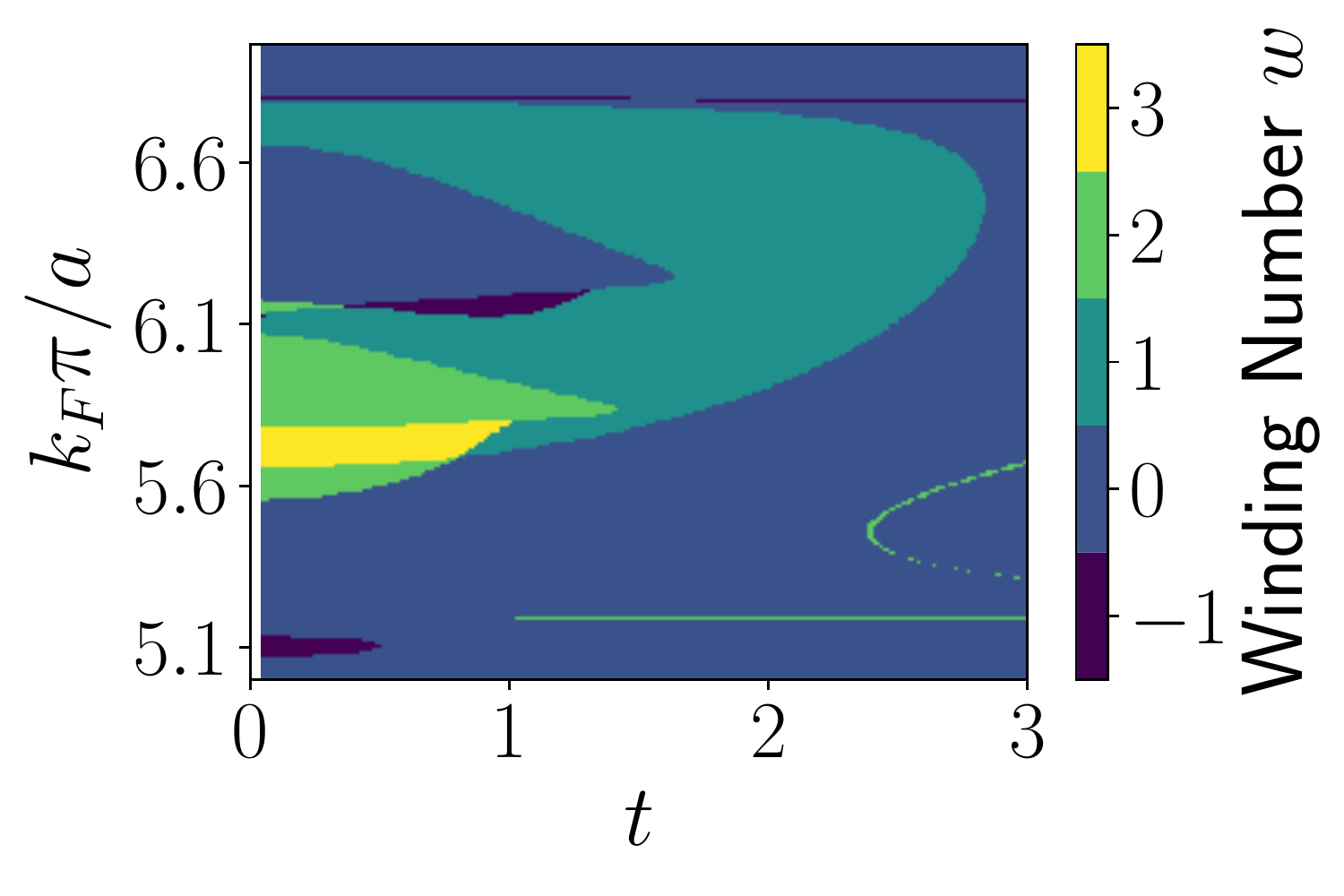}&\includegraphics*[width=0.48\columnwidth]{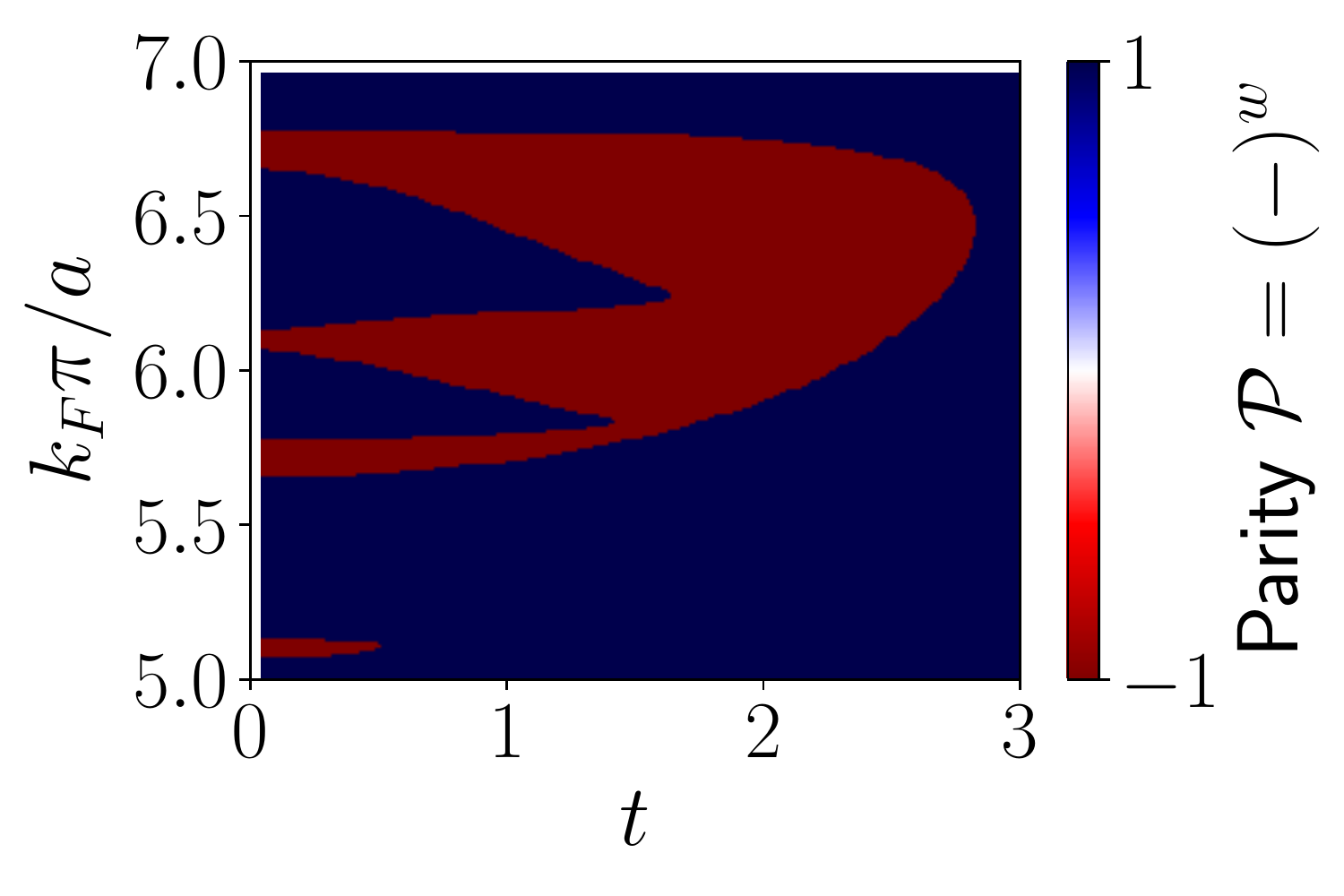}   \\
		\includegraphics*[width=0.48\columnwidth]{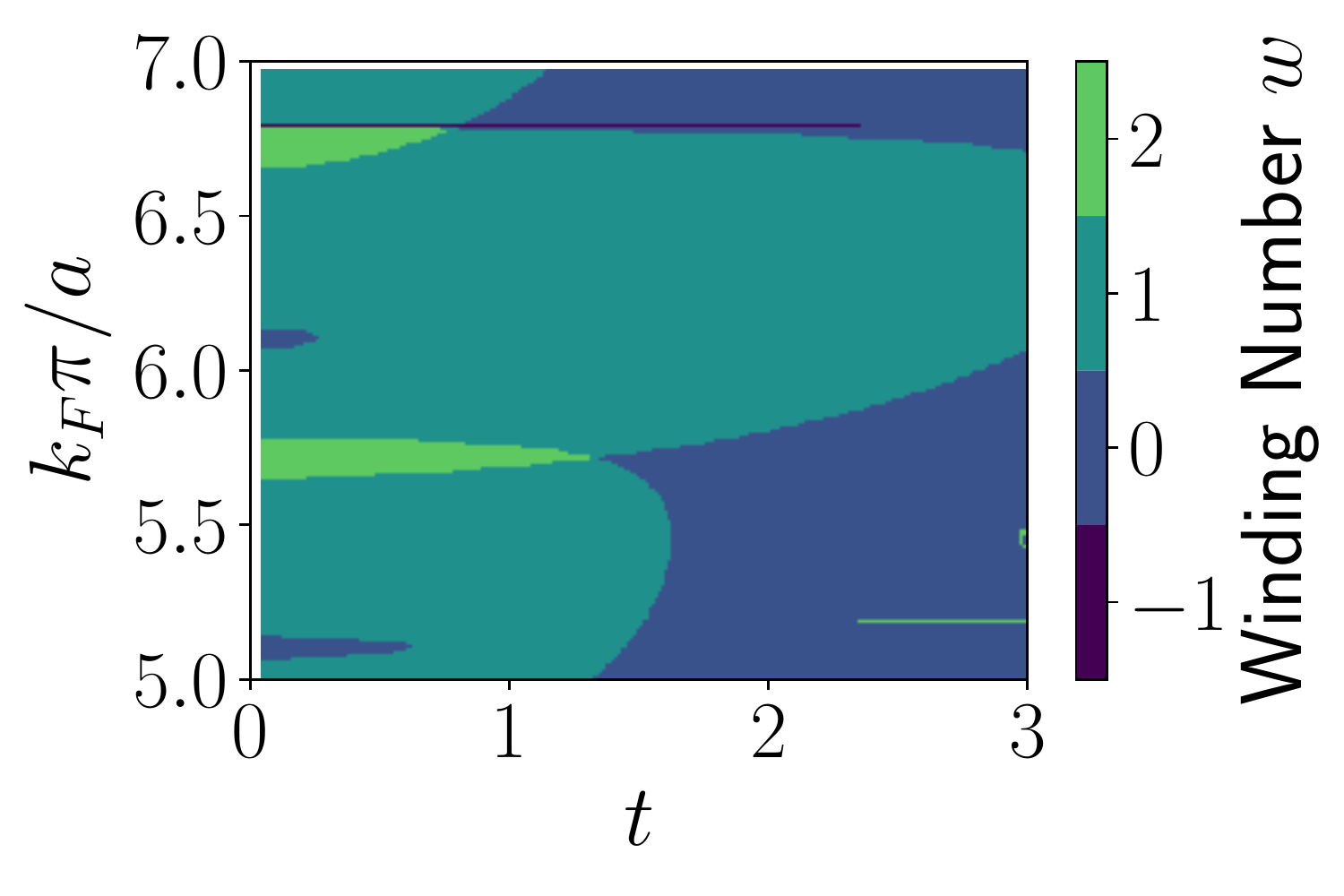} & \includegraphics*[width=0.48\columnwidth]{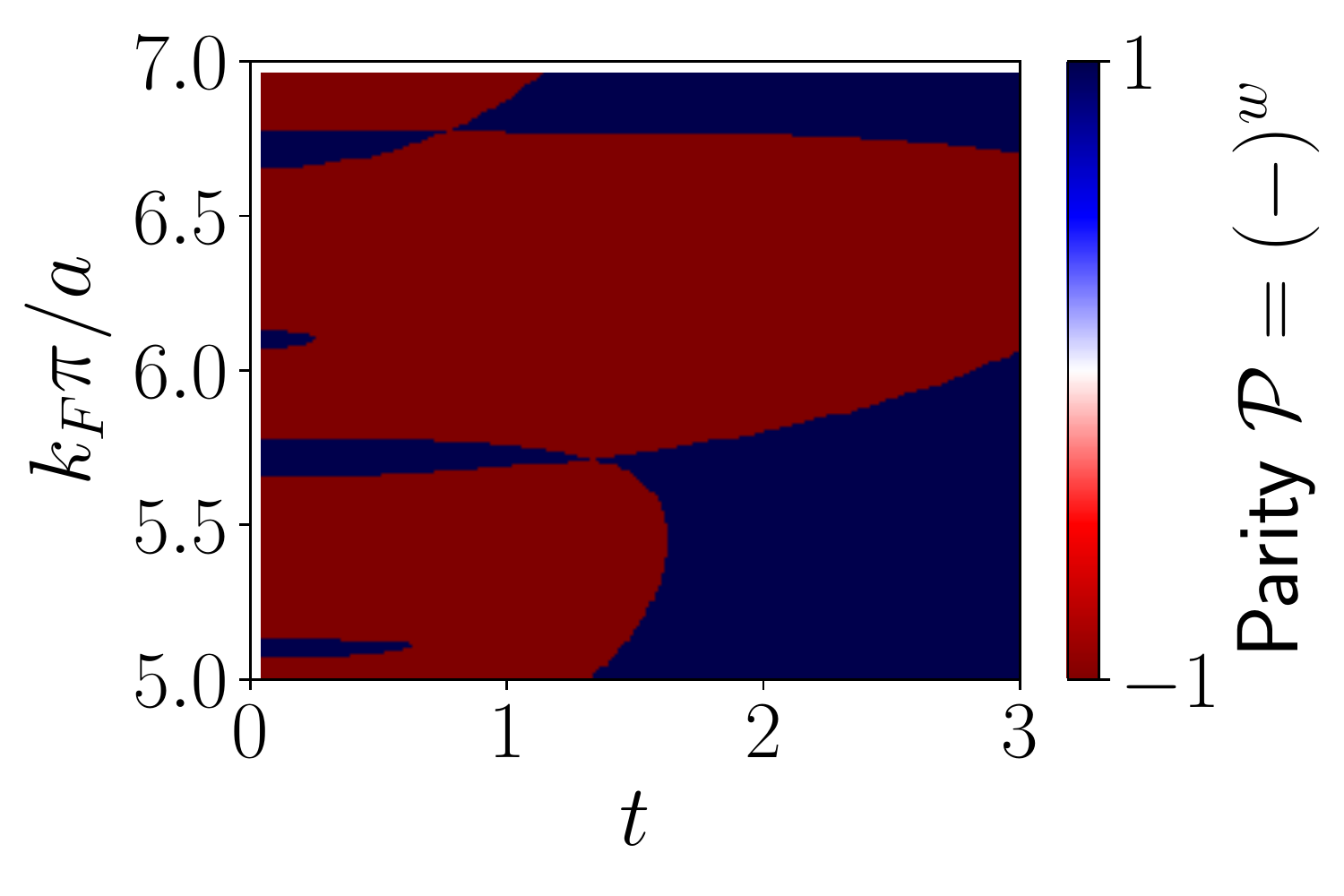} 
\end{tabular}
\caption{(Color on line) Same as in Fig. \ref{fig-phase-diagram1} 
except that we fixed $\alpha=1$ and we are varying $k_F$ instead.  The other parameters are similar to  Fig. \ref{fig-phase-diagram1}.
}
	\label{fig-phase-diagram2}
\end{figure}


Other parameters are important such as the period of the spin helix and the distance between the magnetic atoms.
We analyze the effects of these parameters on the  topological properties in what follows. We plot in Fig. \ref{fig-phase-diagram2} the winding number $w$ as a function of $t$ and $k_F a$, with $J'S=10.5$ (upper panel)  and $J'S=9.8$ (lower panel) and assume $k_h a$ remains constant when varying $a$. For $J'S=10.5$, the wire is initially topologically non-trivial. Therefore at weak $t$ we can get up to three MBS. At very strong hybridization, the strong overlap between the bands destroy their topological character.


\section{Conclusion}
We have considered an array of helical magnetic impurities on a 2D superconducting substrate with two types of orbitals: some localized d-like polarized orbitals which we approximated by classical magnetic moments  and some extended s-like orbitals which form a delocalized band. We have studied the interplay between the 1D Shiba band physics arising from the exchange interaction between the polarized orbitals and the 2D substrate and the delocalized (wire) band on top of the superconductor. 
Both bands can be topological in some parameter space and host MBS at the extremities of the chain: In the dense impurity limit, the magnetic atoms form a ferromagnetic wire above the substrate proximitized by the superconductor similarly to  experiments with semiconducting wires.\cite{Mourik2012,Albrecht2016} In the dilute regime, the magnetic atoms form a Shiba band in the substrate which can be topological.\cite{Pientka2013}  We studied here an intermediate regime where both bands shall be  taken into account. This occurs when the exchange interactions between the magnetic moments and  the electrons in both the 2D substrate and in the delocalized wire band are large and comparable. Possible experimental systems include array of magnetic atoms at intermediate distance \cite{Wiesendanger} or supramolecular assemblies of magnetic organic molecules such as 
porphyrin-based molecular nanowires \cite{Zheng2016} or Mn-based
metalorganic networks.\cite{Giovanelli2014}
In this regime,
  we have derived an effective low-energy Hamiltonian which describes two coupled Kitaev-like Hamiltonian in Eq. \eqref{eq:heff} and analyzed its topological properties. If we assume a perfect planar helix for the initial magnetic chain, and no other inhomogeneities, we found that the low-energy Hamiltonian has an effective TRS symmetry which casts it in the BDI class. We have shown that this effective TRS can be traced back to a magnetic mirror symmetry of the system.\cite{Bernevig2014,NP2014,Li2014}
 If these conditions are satisfied, the system can host multiple Majorana bound states, up to three for a 2D substrate. 
We have numerically computed the phase diagrams of the system depending on the magnetic exchange interactions, the impurity distance and especially the matrix elements between the Shiba and wire band.

When the  magnetic mirror symmetry is broken (thus the effective TRS), we have found that the phase diagram simplifies drastically.
This can typically occur for a non-planar helix or if  some disorder is present either in the substrate or in the chain. Indeed, this automatically makes the coupling between both bands complex and inhomogeneous.
This latter situation should be generic in such complex experimental setup. In that situation, the system enters  the D class and can host one or no MBS at its ends. In particular, when both Shiba and  delocalized wire bands can separately host MBS,  their coupling entails a splitting of the MBS. Therefore, in this intermediate regime of coexistence of the wire and Shiba bands, the system can become non-topological.

\section{Acknowledgements}
We  acknowledge useful discussions with S. Guissart, V. Kaladzhyan, T. Ojanen, and M. Trif.  This work was supported by the French Agence Nationale de la
Recherche through the contract ANR Mistral.
\appendix

\section{Magnetic mirror symmetry} \label{MMS}
The magnetic group symmetry is a combined anti-unitary symmetry composed of a mirror reflection and the usual time reversal symmetry.
Here we introduce
\beq M_T=M_{xz}T,\eeq
 where $M_{xz}$ is the mirror symmetry
with respect to the  $(xz)$ plane (remember $x$ is the axis of the chain and $z$ the direction orthogonal to the substrate) and 
$T=i\sigma^y K$ the TRS operator. 
Since we do not take into account the orbital momentum of the electronic orbitals, the mirror symmetry simply reads as\cite{Bernevig2014,Li2014}
\beq
M_{xz}=i\sigma^y{\cal M}(y\to -y),\eeq
 where ${\cal M}(y\to -y)$ denotes the real space mirror symmetry.
Therefore, if the system is invariant under 
the spatial mirror ${\cal M}(y\to -y)$, $M_T=K$ simply reduces to the complex conjugation and thus to our effective time reversal symmetry. In other words, our effective time-reversal symmetry operator, $\mathcal{T_{\rm eff}}=K$, is the low energy representation of the magnetic mirror symmetry operator $M_T$.


\bibliography{Biblio_Shiba}
\end{document}